\UseRawInputEncoding
\documentclass[final,3p,times]{elsarticle}

\usepackage{amssymb}

\usepackage{booktabs}
\usepackage{amsmath}
\usepackage{multirow}
\usepackage{url}
\usepackage{subcaption}
\usepackage{makecell}
\usepackage{arydshln}
\usepackage{xcolor}
\usepackage{array}
\usepackage{float}
\usepackage{algorithm}
\usepackage[noend]{algpseudocode}
\journal{Speech Communication}

\begin{document}

\begin{frontmatter}



\title{Speaker-Conditioned Phrase Break Prediction for Text-to-Speech with Phoneme-Level Pre-trained Language Model}


\author[inst1]{Dong Yang\corref{cor1}}
\ead{ydqmkkx@gmail.com}
\author[inst1]{Yuki Saito}
\author[inst1]{Takaaki Saeki}
\author[inst2]{Tomoki Koriyama}
\author[inst1]{Wataru Nakata}
\author[inst1]{Detai Xin}
\author[inst1]{Hiroshi Saruwatari}

\cortext[cor1]{Corresponding author.}

\affiliation[inst1]{organization={The University of Tokyo},
            addressline={Hongo}, 
            city={Bunkyo City},
            postcode={113-8656}, 
            state={Tokyo},
            country={Japan}}

\affiliation[inst2]{organization={CyberAgent, Inc.},
            addressline={Shibuya}, 
            city={Shibuya City},
            postcode={150-6101}, 
            state={Tokyo},
            country={Japan}}

\begin{abstract}

This paper advances phrase break prediction (also known as phrasing) in multi-speaker text-to-speech (TTS) systems. We integrate speaker-specific features by leveraging speaker embeddings to enhance the performance of the phrasing model. We further demonstrate that these speaker embeddings can capture speaker-related characteristics solely from the phrasing task. Besides, we explore the potential of pre-trained speaker embeddings for unseen speakers through a few-shot adaptation method. Furthermore, we pioneer the application of phoneme-level pre-trained language models to this TTS front-end task, which significantly boosts the accuracy of the phrasing model. Our methods are rigorously assessed through both objective and subjective evaluations, demonstrating their effectiveness.

\end{abstract}



\end{frontmatter}


\section{Introduction}
\label{sec: introduction}

Text-to-speech (TTS) systems~\cite{tan2021survey0, khanam2022survey1} aim to generate intelligible, natural-sounding artificial speech from a given text with a specific speaker's voice. Typically, TTS systems comprise two primary components: the front-end for text processing and the back-end for speech synthesis. Recent advancements in neural end-to-end generative TTS models (acoustic models~\cite{ping18deepvoice3, shen18tacotron2, ren21fastspeech2, popov21gradtts, kim21vits, kong23vits2, matcha-tts} and vocoders~\cite{oord16wavenet, kumar19melgan, kong20hifigan, kong21diffwave, vocos}), have enabled synthetic speech to closely resemble natural human voice and effectively capture the speaker's acoustic features. Consequently, enhancements to the front-end of TTS systems are receiving increased attention, intending to refine synthetic speech's naturalness by exploiting richer linguistic and semantic information and capturing more human speech patterns. Notably, pause insertion is indispensable to improve synthetic speech's naturalness, prosody, and rhythm, especially in long-form TTS synthesis~\cite{klimnov17longformpause}. 

Inserting pauses in speech is a natural phenomenon in human communication, often done to enhance expressiveness or to take a breath. In TTS, pauses can broadly be classified into two types: \textit{respiratory pauses (RPs)}~\cite{bailly12respause, klimnov17longformpause} and \textit{punctuation-indicated pauses (PIPs)}. RPs are inserted at word transitions without punctuation, whereas PIPs occur at punctuation marks, aligned with textual indications. Since speakers commonly insert PIPs at punctuation marks during reading, predicting the position of RPs for a given text is more challenging for TTS systems, termed \textit{phrase break prediction} or \textit{phrasing}. Many TTS systems~\cite{abbas22cauliflow, xue22paratts} have integrated a predictive model for phrasing (referred to as a phrasing model) into their front-end, which utilizes word representations derived from a pre-trained language model (PLM). Since Transformer~\cite{vaswani17attention} encoder-based PLMs with self-supervised learning (SSL), such as BERT~\cite{devlin19bert}, have demonstrated superior word representations across various natural language processing (NLP) tasks, they become increasingly adopted in TTS models~\cite{addtts0, addtts1, addtts2, addtts3} and phrasing models~\cite{futamata21japanesepause, lee23crosslingualpause, phrasing-cited}. 

From our empirical observations, we identify two key challenges in current phrasing models:

\textbf{Challenges 1.} In speech corpora, RPs are sparsely distributed, necessitating the use of large datasets to develop robust phrasing models. However, large speech corpora typically involve multiple speakers. Although RP insertion follows some latent rules~\cite{klimnov17longformpause}, different speakers have different styles of inserting RPs. 
The phrasing models that ignore speaker identity are forced to learn an ``average'' phrasing behavior, treating individual stylistic variations as noise, which inherently limits their predictive accuracy.

\textbf{Challenges 2.} While mainstream PLMs, such as BERT\textsubscript{BASE}, have enhanced the performance of phrasing models, they typically generate representations at the subword level. Their effectiveness in phrasing tasks suggests that RP insertion is correlated with syntactic and semantic features. However, RP insertion is fundamentally a spoken-language phenomenon, governed by acoustic and articulatory mechanisms, whereas subword tokens are merely abstract textual units that lack direct information about acoustic features.

To tackle these two challenges, we put forward two main hypotheses and propose several methods not only to verify them but also to explore further implications and applications:

\textbf{Hypothesis 1.} The performance limitations of the phrasing model on multi-speaker datasets stem from the absence of speaker-specific features.

\textbf{Method 1.} Incorporating speaker embeddings into the phrasing model. In multi-speaker TTS models, speaker embeddings are widely used to capture and control the timbre and prosody characteristics of speakers, represented as fixed-length vectors. We propose inserting the speaker embedding between the encoder and decoder of the phrasing model to enable the model to learn the speaker-specific features. Besides, in TTS models, speaker embeddings can be either randomly initialized or derived from a pre-trained speaker verification model (PSVM). The former approach is restricted to the speakers included in the training set, referred to as \textit{seen speakers}. The latter can generalize to \textit{unseen speakers} which were not present during training. These methods present an opportunity to investigate whether the prosody and rhythm features encoded in PSVM embeddings can be effectively leveraged in phrasing tasks.

\textbf{Hypothesis 2.} As the fundamental units of speech, phonemes have proven to be highly effective inputs in TTS synthesis models~\cite{zhang22mpbert, kong23vits2}. Consequently, phoneme representations, which inherently encode richer acoustic attributes than subword representations, could potentially enhance the predictive accuracy of phrasing models.

\textbf{Method 2.} Applying phoneme-level PLMs in phrasing models. Within the BERT framework, PLMs that use phoneme tokens as input have been developed: mixed-phoneme BERT~\cite{zhang22mpbert} (MP BERT) and phoneme-level BERT~\cite{li23plbert} (PL BERT). They have primarily served as the phoneme encoder of TTS models, where the included semantic information contributes to the naturalness of synthesized speech. However, their potential applications in other domains remain unexplored. The unique design of pre-training objectives, which enrich generated representations with both word and phoneme information, suggests that these phoneme-level PLMs could be utilized in phrasing tasks to validate our hypothesis.

In our previous work~\cite{yang23pause}, we proposed a speaker-conditioned phrasing model and a duration-aware pause insertion model. In this paper, we focus on the speaker-conditioned phrasing model. We first made several critical improvements:
\begin{itemize}
    \item \textbf{Dataset}: We construct the phrasing dataset based on the LibriTTS-R corpus~\cite{koizumi23librittsr}, facilitating pause detection and speech synthesis due to its higher sound quality (Section~\ref{sec: dataset}).
    
    \item \textbf{Model and training}: To address the overfitting problem and instability during training, we improve the architecture of both the baseline and proposed models and optimize the training configurations (Section~\ref{sec: proposed}, Section~\ref{sec: configs baseline}, and Section~\ref{sec: configs training}).
    
    \item \textbf{Synthesis quality}: In the subjective evaluations of our previous work, we found FastSpeech 2~\cite{ren21fastspeech2} with HiFi-GAN~\cite{kong20hifigan} struggled to synthesize breath sounds, which are important components of pauses and significantly affect the listeners' perception. Therefore, we instead use both VITS~\cite{kim21vits} and Matcha-TTS~\cite{matcha-tts} (see~\ref{appendix a} for more details) as the backbone TTS models, which offer natural breath synthesis and benefit the evaluation.
\end{itemize}

Then, we extend our previous work with the following contributions:
\begin{itemize}
    \item \textbf{Exploration of PSVM embeddings}: We incorporate speaker embeddings from various PSVMs into our proposed phrasing models. This exploration aims to determine whether the prosody and rhythm features encoded in these embeddings can be leveraged in phrasing tasks (Section~\ref{sec: eval spk}).
    
    \item \textbf{Leverage of various PLMs}: We experiment with advanced subword-level PLMs in our proposed model and demonstrate improved phrasing results with some of them. Besides, we apply two phoneme-level PLMs to our proposed model. The phoneme-level PLMs outperform subword-level PLMs, confirming our hypothesis that phoneme features can improve the accuracy of the phrasing task. This is also the first application of phoneme-level PLMs in a TTS front-end task (Section~\ref{sec: eval plm}).
    
    \item \textbf{Few-shot adaptation for unseen speakers}: For speakers not present in the training set, we demonstrate that PSVM embeddings can equip the phrasing model with few-shot learning capabilities. Using several utterances enables the model to outperform the baseline model without additional training (Section~\ref{sec: eval few-shot}). 

    \item \textbf{Analysis on speaker embeddings}: We provide a detailed analysis showing that the speaker embeddings in our proposed phrasing model capture meaningful speaker-related characteristics, despite being trained solely on the phrasing task without any additional objectives. This finding provides support for our use of PSVM embeddings and the proposed few-shot adaptation method (Section~\ref{sec: discussion}).
\end{itemize}

\section{Background}
\subsection{Phrasing}\label{sec: phrasing}
In the preprocessing phase of TTS training, the text in the training set is aligned with the utterance or directly obtained through speech recognition. In this process, pauses in the utterances are detected and annotated as pause marks in the text, which enables the TTS model to learn to generate pauses at these marks during training. However, during inference, the original text can only be used to obtain PIPs by inserting pause marks at punctuation without generating RPs. Therefore, the phrasing model is employed during inference to predict appropriate RP positions based on the text, and pause marks are then inserted at these predictive positions.

Traditional TTS systems mainly focused on short-form speech, which contains few pauses, making RPs rare and less necessary. However, with the increasing application of long-form TTS, the phrasing model has become an essential module of TTS systems and follows advances in NLP technology. Initially, since people's behavior in inserting RPs is guided by some grammatical rules, lexical and syntactic features like part-of-speech tags~\cite{chiche22postag} were utilized in phrasing tasks. Some machine learning models, such as tree algorithms~\cite{tree1, tree2, tree3, tree4, tree5, klimnov17longformpause} and hidden Markov models~\cite{hmm1, hmm2, hmm3}, were used to process these features and predict RPs. Subsequently, the advent of word2vec~\cite{mikolov13word2vec} and the efficacy of recurrent neural networks (RNNs)~\cite{hochreiter97lstm, sak14lstmp} in text processing further enhanced phrasing accuracy. 

In recent years, PLMs with Transformer encoders as the backbone have demonstrated powerful performance in textual feature extraction. Notably, Futamata et al.~\cite{futamata21japanesepause} have incorporated features from Japanese BERT in a Japanese phrasing task, and Abbas et al~\cite{abbas22cauliflow} have implemented word-level BERT embeddings as inputs in a phrasing model. Consequently, the combination of BERT and bidirectional long short-term memory (BiLSTM)~\cite{graves05bilstm} has become a generalized approach for phrasing tasks.

\subsection{Speaker embeddings}
Speaker embedding, originally developed for speaker recognition, is extracted from a speech utterance and represents the speaker's identity in a fixed-dimensional vector~\cite{residualinfo, speakerembed}. Speaker verification, a subtask of speaker recognition, aims to verify whether an utterance is pronounced by a target speaker based on its recorded utterances~\cite{speakerrecognition}. PSVMs output a speaker embedding for each input utterance and are pre-trained with a large number of utterances from various speakers. Recent advances have enabled these models to be trained end-to-end, incorporating additive angular margin loss~\cite{aam} to minimize the distance between embeddings from the same speaker and maximize the distance between those from different speakers. After training, cosine similarity is used to measure the distance between speaker embeddings. Notably, Voxceleb 1~\cite{nagrani17voxceleb} and Voxceleb 2~\cite{chung18voxceleb2} datasets are commonly involved in the training, with the equal error rate (EER) on the Voxceleb 1 clean test set serving as a key metric for evaluating the models' performance.

Deep Voice 2~\cite{gibiansky17deepvoice2} innovatively applied speaker embeddings to multi-speaker TTS, with each embedding corresponding to one speaker. These embeddings were initialized randomly and trained jointly with the TTS model, effectively capturing speakers' features and facilitating natural TTS for various speakers. This method has been widely adopted in multi-speaker TTS. Building on this, Chen et al.~\cite{chen19fewshot} extended the utility of a pre-trained multi-speaker TTS model for speakers not present in the training set. By fine-tuning the model with just a few utterances from these unseen speakers, they achieved effective few-shot learning for speech synthesis. Furthermore, Jia et al.~\cite{jia18speakerembedtts} implemented a transfer learning approach by integrating speaker embeddings from a PSVM into a TTS model. During training, the speaker embeddings were frozen, enabling the TTS model to generalize to unseen speakers during inference. This approach eliminated the need to update the model's parameters and achieved excellent zero-shot learning results. Its success demonstrates that the speaker embeddings from the PSVM contain speakers' features in timbre and prosody, which can be effectively utilized by TTS models.

\subsection{Phoneme-level PLMs}\label{sec: pl plms}
TTS models that use phoneme sequences as input typically incorporate a phoneme encoder to capture the relationships among phonemes. Given that BERT is powerful in extracting contextual relationships due to its bidirectional attention mechanism and the masked language modeling (MLM) objective during pre-training, employing a BERT-based phoneme encoder pre-trained with phoneme sequences proves effective in TTS. This approach enriches the TTS model with semantic features and then enhances the naturalness of the synthesized speech. Consequently, within the BERT framework, PLMs that use phoneme tokens as input have been developed: Mixed-Phoneme BERT~\cite{zhang22mpbert} (MP BERT) and Phoneme-Level BERT~\cite{li23plbert} (PL BERT).

\subsubsection{MP BERT}
Phoneme-level PLMs suffer from a notably small phoneme dictionary size, approximately 200 entries, in contrast to the more extensive word or subword dictionaries. This limitation restricts the models' representation capacity and effectiveness in conveying contextual semantic information~\cite{ding19vocab0, Kharitonov21vocab1, Mielke21vocab2}. To address this issue, MP BERT introduces ``sup-phoneme'' tokens as an auxiliary representation. By applying byte-pair encoding (BPE) learning~\cite{sennrich16bpe}, the phoneme sequences are encoded into the sup-phoneme sequences, where each sup-phoneme refers to a group of neighboring phonemes. The sup-phoneme tokens are up-sampled and aligned with the corresponding phoneme tokens and position IDs before being fed into the vanilla BERT~\cite{devlin19bert} architecture. During the pre-training phase, MP BERT adopts two MLM objectives: predicting masked phoneme and sup-phoneme tokens. 

\subsubsection{PL BERT}
Li et al.~\cite{li23plbert} pinpointed several limitations of MP BERT, particularly the lack of assurance that sup-phoneme tokens carry as much linguistic information as graphemes and incorporating sup-phoneme tokens diminishes training and inference speeds. To overcome these drawbacks, they have proposed PL BERT, which introduces two objectives for pre-training: MLM for phoneme tokens and phoneme-to-grapheme (P2G) prediction. PL BERT takes phoneme tokens and position IDs as input, leveraging the ALBERT~\cite{lan20albert} architecture as its backbone. During pre-training, PL BERT predicts masked phoneme tokens and the corresponding grapheme tokens for each phoneme. The P2G prediction enriches phoneme representations with word semantics.

\section{Proposed phrasing model}\label{sec: proposed}

\subsection{Model architecture}

\begin{figure*}
    \centering
    \includegraphics[width=0.8\linewidth, clip]{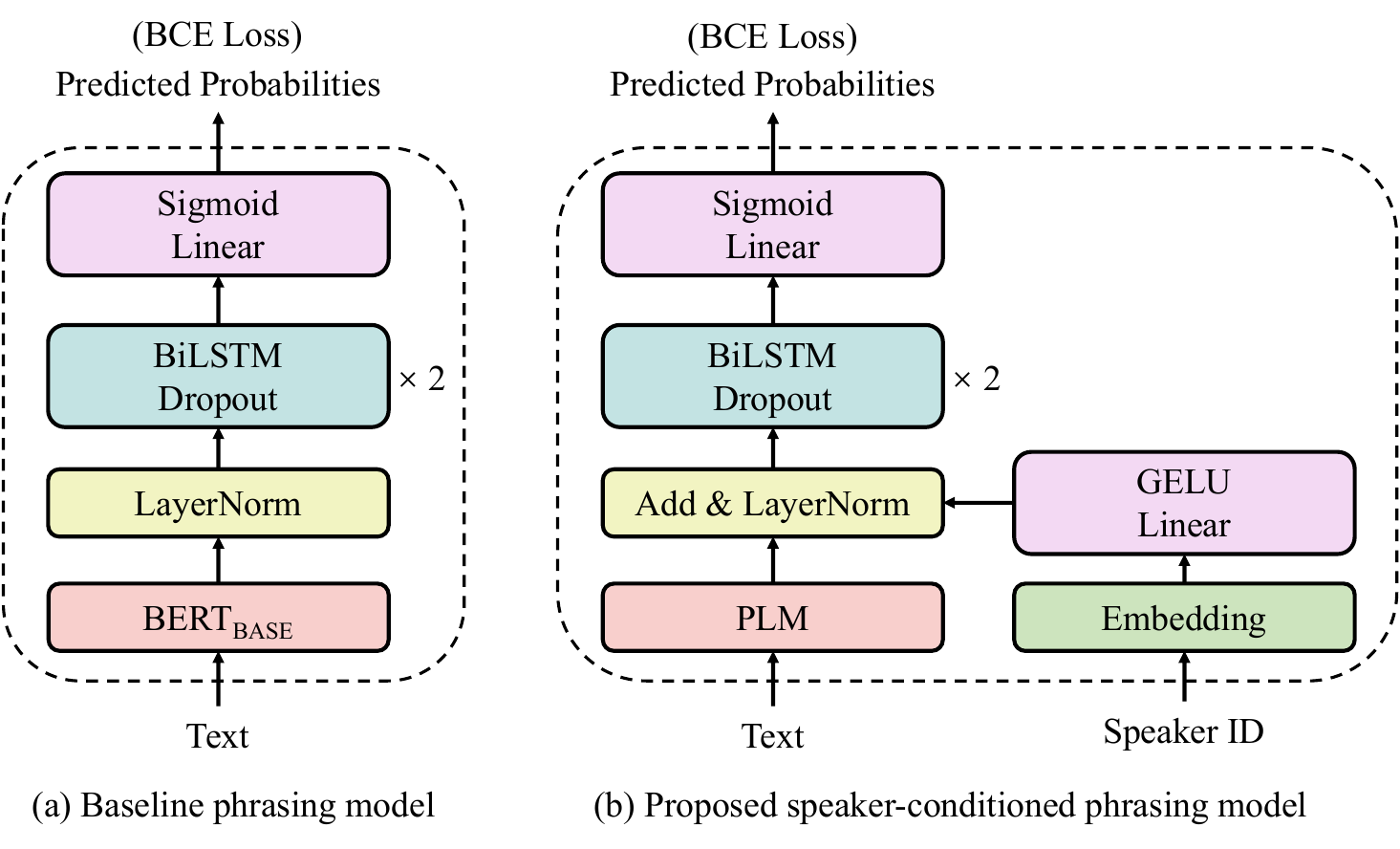}
    \caption{Architecture of phrasing models.}
    \label{fig: architecture}
\end{figure*}

\begin{figure*}
    \centering
    \includegraphics[width=0.9\linewidth, clip]{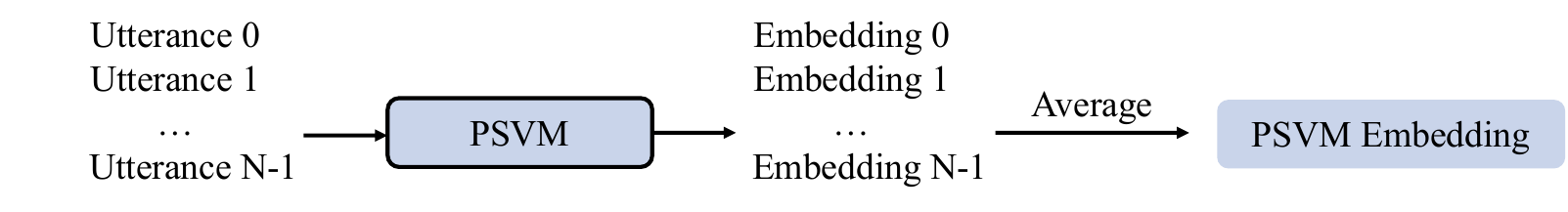}
    \caption{PSVM embedding extraction pipeline for a single speaker.}
    \label{fig: embedding}
\end{figure*}

As illustrated in Fig.~\ref{fig: architecture} (b), the proposed speaker-conditioned phrasing model is improved from our previous work~\cite{yang23pause}. Similar to conventional phrasing models~\cite{klimnov17longformpause, futamata21japanesepause, abbas22cauliflow}, it adopts an encoder-decoder architecture, where the encoder is a PLM that provides linguistic features. To incorporate speaker information, a speaker embedding layer converts the input speaker ID into an embedding, which is linearly projected to match the PLM's output size, using the GELU~\cite{gelu} function for non-linearity. The decoder contains two BiLSTM layers to enhance sequence processing. It leverages both linguistic and speaker-specific features to predict the probability of an RP occurrence with the Sigmoid function at the output layer. Due to the sparse distribution of RPs, dropout and layer normalization are applied to mitigate overfitting and facilitate convergence, respectively.

\subsection{Speaker embedding}
For the speaker embedding layer in our proposed model, we experimented with both randomly initialized embeddings and speaker embeddings extracted using several PSVMs, inspired by \cite{jia18speakerembedtts}. As illustrated in Fig.~\ref{fig: embedding}, each PSVM generates a fixed-dimension embedding for a given utterance. Let $\mathcal{M}$ denote the set of speaker IDs in the training set. For the speaker $m \in \mathcal{M}$ with $N_m$ training utterances, we apply a PSVM to each utterance and obtain $N_m$ utterance-level embeddings, denoted as $\{\mathbf{e}_{m,n}\}_{n=0}^{N_m-1}$. The speaker-level embedding $\mathbf{e}_m$ is then computed as the average of these embeddings:

\begin{equation}
    \mathbf{e}_m = \frac{1}{N_m} \sum_{n=0}^{N_m-1} \mathbf{e}_{m,n}.
\end{equation}
Using this approach, we obtained the PSVM embeddings of seen speakers $\{\mathbf{e}_m\}_{m \in \mathcal{M}}$, which were used to initialize the embedding layer in Fig.~\ref{fig: architecture} (b); that is, the embeddings were stored as the layer's weights, indexed by the corresponding speaker IDs. During training and inference, only the speaker IDs were required as input and the utterances were not used. The effect of these PSVM embeddings was explored in Section~\ref{sec: eval spk}.

\subsection{PLM}
Although BERT\textsubscript{BASE} has been utilized in previous studies on phrasing tasks, more advanced subword-level PLMs that surpass its capabilities in the NLP field have not been explored. Furthermore, as discussed in Section~\ref{sec: introduction}, we hypothesize that incorporating phoneme-level features could enhance the predictive accuracy of phrasing tasks. Therefore, we conducted experiments with various subword-level and phoneme-level PLMs in Section~\ref{sec: eval plm}. 

To argue that phoneme-level representations ($X_{\text{phoneme}}$) are more effective than subword-level representations ($X_{\text{subword}}$) in phrasing tasks, we consider the mutual information ($I(\cdot)$) between the input and RP labels, aiming to demonstrate that
\begin{equation}
    I(X_{\text{phoneme}}; Y_{\text{word}}) > I(X_{\text{subword}}; Y_{\text{word}}). 
\end{equation}
RP labels ($Y_{\text{word}}$) form binary sequences, where each element indicates whether an RP follows a word. We denote the entropy as $H(\cdot)$, according to 
\begin{equation}
    I(X;Y_{\text{word}})=H(Y_{\text{word}}) - H(Y_{\text{word}} | X),
\end{equation}
when $Y_{\text{word}}$ is identical for all phrasing models, the inequality holds if we can verify that 
\begin{equation}
    H(Y_{\text{word}} | X_{\text{phoneme}})<H(Y_{\text{word}} | X_{\text{subword}}).
\end{equation}
This indicates that the uncertainty associated with RP insertion would be lower when the phrasing model is conditioned on phoneme representations, suggesting that phrasing models with phoneme-level PLMs are expected to achieve higher prediction accuracy than those with subword-level PLMs. A crucial aspect of the verification is ensuring the consistency of the label space $Y_{\text{word}}$, which we achieve by evaluating the prediction accuracy strictly at the word level, as detailed in Section~\ref{sec: dataset}.

\subsection{Embedding adapter}
\label{sec: proposed adapter}

\begin{figure}
    \centering
    \includegraphics[width=1.0\linewidth, clip]{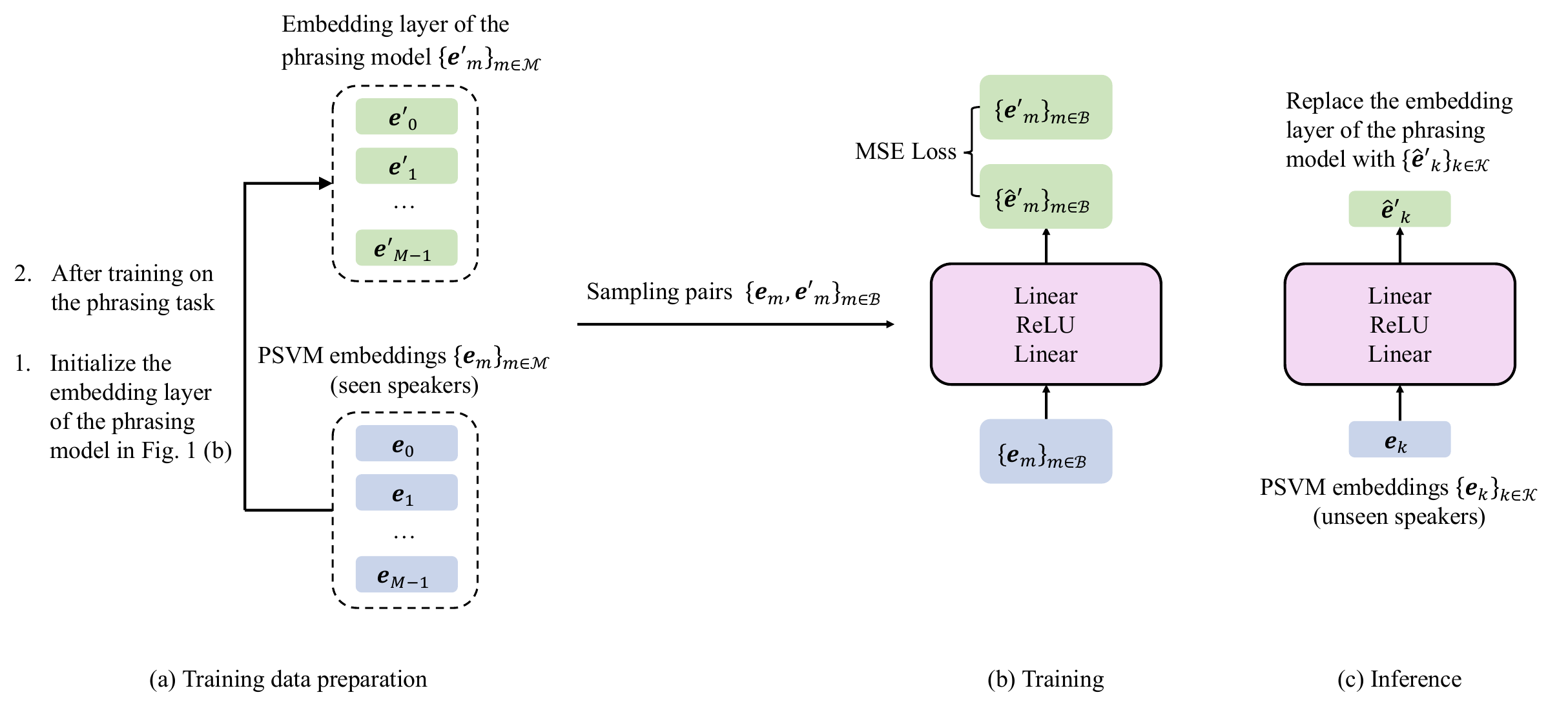}
    \caption{Architecture, training, and inference of the embedding adapter.}
    \label{fig: adapter}
\end{figure}

In Section~\ref{sec: eval few-shot}, we explored the potential of PSVM embeddings to handle unseen speakers using a few-shot adaptation method. We denote the set of unseen speaker IDs as $\mathcal{K}$ and their PSVM embeddings as $\{\mathbf{e}_k\}_{k \in \mathcal{K}}$.
There were two scenarios:

\begin{itemize}
    \item \textbf{Fixed speaker embedding layer:} When the phrasing models are pre-trained with a fixed speaker embedding layer, the embedding layer is initialized with $\{\mathbf{e}_m\}_{m \in \mathcal{M}}$ and not updated. Consequently, the models can directly handle unseen speakers by replacing the embedding layer with $\{\mathbf{e}_k\}_{k \in \mathcal{K}}$.

    \item \textbf{Trainable speaker embedding layer:} When the phrasing models are pre-trained with a trainable speaker embedding layer, the speaker embeddings $\{\mathbf{e}_m\}_{m \in \mathcal{M}}$ are updated as $\{\mathbf{e}'_m\}_{m \in \mathcal{M}}$ during training. In this case, the models cannot generalize to unseen speakers directly by using $\{\mathbf{e}_k\}_{k \in \mathcal{K}}$ as the embedding layer.
\end{itemize}

For the second scenario, since $\{\mathbf{e}_m\}_{m \in \mathcal{M}}$ and $\{\mathbf{e}'_m\}_{m \in \mathcal{M}}$ are pairwise distinct, we assume that there exists an injective mapping from $\{\mathbf{e}_m\}_{m \in \mathcal{M}}$ to $\{\mathbf{e}'_m\}_{m \in \mathcal{M}}$. To approximate this mapping, we propose and further train simple embedding adapters for each pre-trained phrasing model: $\boldsymbol{f_\theta}: \mathbb{R}^d \rightarrow \mathbb{R}^d$, where $d$ denotes the dimension of the embeddings. As shown in Fig.~\ref{fig: adapter}, the embedding adapter consists of two linear layers with a ReLU activation function applied between them. After training the adapter, it can map the PSVM embeddings of unseen speakers, $\{\mathbf{e}_k\}_{k \in \mathcal{K}}$, to their adapted counterparts, $\{\hat{\mathbf{e}}'_k\}_{k \in \mathcal{K}}$. By replacing the embedding layer in the pre-trained phrasing model with $\{\hat{\mathbf{e}}'_k\}_{k \in \mathcal{K}}$, the pre-trained phrasing model can then be applied to handle unseen speakers. For further clarity, the detailed algorithm is presented in~\ref{appendix b}.

\section{Experimental configurations}
\label{sec: configs}

\subsection{Dataset construction}
\label{sec: dataset}

\begin{table*}[t]
  \caption{Statistics of the datasets.}
  \label{tab: statistics}
  \centering
  \small
  \begin{tabular}{l r r r r}
    \toprule
    Set & Utterances & Speakers & Words & RPs\\
    \midrule
    Training & 283,769 & 2,308 & 5,220,392 & 159,047\\
    Validation-seen & 35,000 & 2,239 & 640,475 & 19,554\\
    Test-seen & 35,000 & 2,240 & 644,110 & 19,700\\
    Validation-unseen & 4,700 & 94 & 87,350 & 1,690\\
    Test-unseen & 13,000 & 94 & 226,114 & 3,889\\
    \bottomrule
  \end{tabular}
\end{table*}

We constructed a large-scale phrasing dataset from the LibriTTS-R corpus, a multi-speaker speech corpus derived from audiobooks and enhanced with speech restoration. It includes numerous long-form utterances with multiple pauses voiced by 2,456 speakers. The speech in the corpus is clean and high-quality, containing pauses mainly made up of silence, breath, and tongue clicks~\cite{yang24breath}, rendering it perfectly suited for the phrasing task. Among the subsets of the LibriTTS-R corpus, ``train-100-clean'', ``train-360-clean'', and ``train-500-other'' were segmented into training, validation-seen, and test-seen sets in an approximately 8:1:1 ratio. The speakers in the validation-seen and test-seen sets are seen speakers. ``dev-clean'', ``dev-other'', ``test-clean'', and ``test-other'' subsets were divided into validation-unseen and test-unseen sets, containing unseen speakers. Specifically, each speaker contributed 50 randomly selected utterances to the validation-unseen set for extracting PSVM embeddings, while the remaining utterances were allocated to the test-unseen set. The validation-seen and test-seen sets were utilized in Section~\ref{sec: seen}, and validation-unseen and test-unseen sets were used in Section~\ref{sec: unseen}. Detailed statistics are provided in Table~\ref{tab: statistics}.

\begin{figure*}
    \centering
    \includegraphics[width=1.0\linewidth, clip]{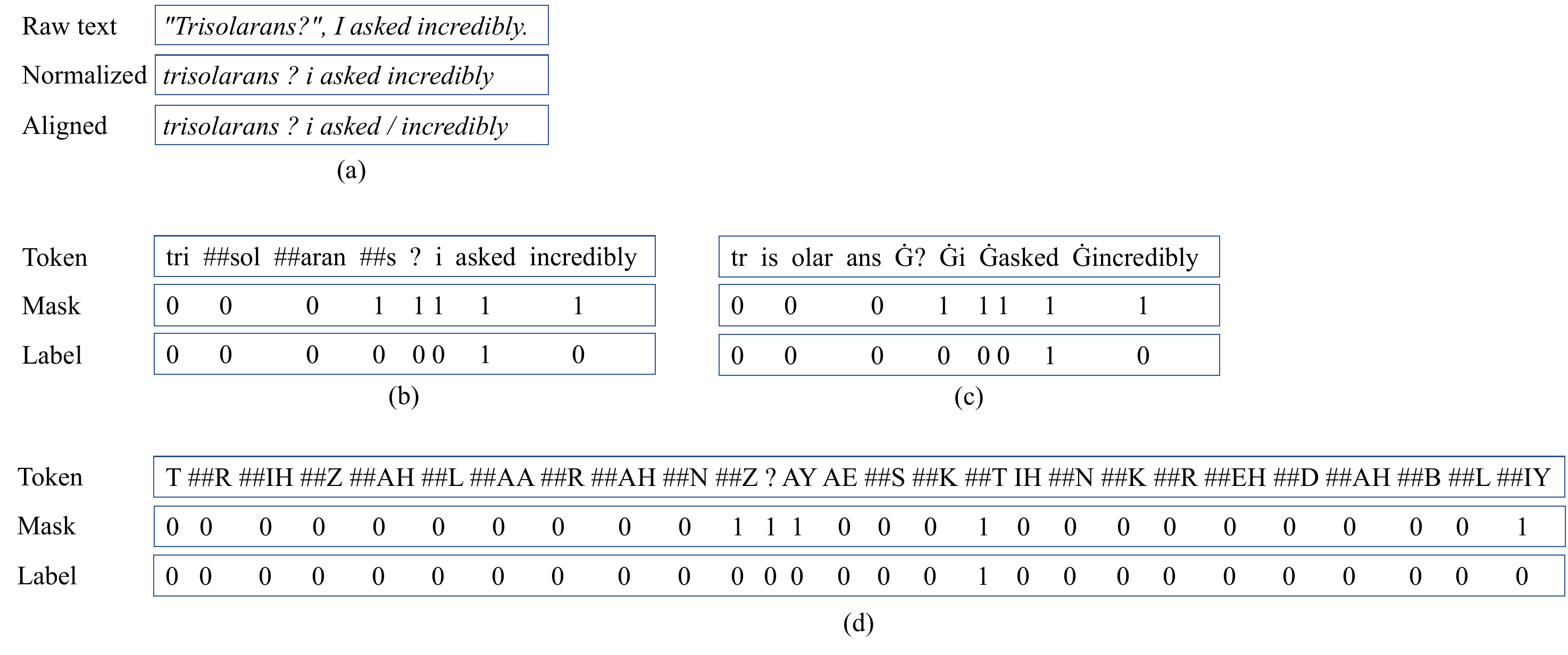}
    \caption{Example of text pre-processing, tokenization, and label generation. The text in the example was not used in the experiments.
    (a) Text pre-processing. The aligned text is stored in the dataset. /: RP mark.
    (b) Using WordPiece~\cite{schuster12wordpiece, devlin19bert} subword tokenizer. \#\#: prefix for continuing subword.
    (c) Using BPE~\cite{sennrich16bpe} subword tokenizer. Ġ: prefix indicating a whitespace.
    (d) Using phoneme tokenizer~\cite{raphe}. \#\#: prefix for continuing phoneme within a word.}
    \label{fig: example}
\end{figure*}

During pre-processing, we normalized the text and converted all characters to lowercase. Montreal Forced Aligner (MFA)~\cite{mcauliffe17mfa} was utilized to align text and speech, identifying pauses with their duration. Then, we further aligned the detected pauses with punctuation marks in the text. We preserved only the first punctuation in sequences of consecutive punctuation and removed punctuation at both the start and end of utterances to ensure more accurate alignment. Pauses exceeding 50 ms at word transitions lacking punctuation were regarded as RPs. As demonstrated in Fig.~\ref{fig: example}, RP marks within the text were used solely for label generation but not fed into the phrasing model. Specifically, the label ``1'' indicates the presence of an RP after the respective token. Given that a word may be split into different tokens by different tokenizers, we assigned the ``0'' mask to all non-final tokens derived from a single word. Only tokens with the ``1'' mask were considered when calculating losses during training and metrics during evaluation. This approach establishes the word as the fundamental unit for phrasing tasks, thereby ensuring fair comparison across various tokenization methods.

\subsection{Baseline settings}
\label{sec: configs baseline}
The baseline phrasing model is shown in Fig.~\ref{fig: architecture} (a). In contrast to our proposed model, it excludes the speaker-related modules and utilizes BERT\textsubscript{BASE}. This setup aligns with conventional phrasing models~\cite{klimnov17longformpause, futamata21japanesepause, abbas22cauliflow} and facilitates a direct comparison with our proposed model. Additionally, when we experiment with the proposed model to explore the impact of different speaker embeddings and PLMs, BERT\textsubscript{BASE} serves as the baseline PLM.

\subsection{Model configurations}
For the subword-level PLM, we tested both BASE and LARGE variants of BERT, XLNET~\cite{yang20xlnet}, RoBERTa~\cite{liu19roberta}, ALBERT~\cite{lan20albert}, and DeBERTaV3~\cite{he23debertav3}. For the phoneme-level PLM, we evaluated MP BERT and PL BERT, which were implemented by ourselves. Resource links are provided in~\ref{appendix c}. 

For the PSVMs, we used officially pre-trained ECAPA-TDNN~\cite{desplanques20ecapa}, ResNet-TDNN~\cite{villalba20resnet}, SpeakerNet~\cite{koluguri20speakernet}, and Titanet\textsubscript{LARGE}~\cite{koluguri22titanet}. The impact of the training data mismatch between the PSVMs and the phrasing models is further explored in~\ref{appendix d}. We experimented with both frozen embeddings, where the speaker embedding layer was fixed, and trainable embeddings, where the parameters of the speaker embedding layer were updated. The randomly initialized speaker embeddings were initialized using Xavier initialization~\cite{glorot10xavier}.

For the phrasing models, each BiLSTM layer in the decoder was configured with a hidden size equal to half the output size of the PLM. The dropout rate was set to 0.5. For the embedding adapter, the first linear layer expanded the dimension to 1024, and the second reduced it to its original size.

\subsection{Evaluation metrics}

\subsubsection{Objective evaluation}
Since the output of the phrasing models is a predictive probability between 0 and 1, a threshold must be chosen to determine whether an RP should be inserted. Furthermore, because it is preferable to skip RPs rather than insert them in inappropriate places~\cite{klimnov17longformpause}, we used F\textsubscript{0.5} score both for threshold selection on the validation-seen set and as the primary objective evaluation metric on the test-seen/-unseen sets:

\begin{align}
    \mathrm{F}_{0.5} &= (1 + 0.5^{2}) \frac{\mathrm{Precision} \times \mathrm{Recall}}{0.5^{2} \times \mathrm{Precision} + \mathrm{Recall}}
    \label{eq:threshold}
\end{align}

\subsubsection{Subjective evaluation}
In addition to objective evaluations focusing on predictive accuracy, subjective evaluations are essential to assess the impact of different phrasing models on the perceived naturalness of synthetic speech. For the subjective evaluations, we chose VITS and Matcha-TTS as the TTS models due to their high-quality generated speech. Sampled transcripts from the test-seen/-unseen sets were phrased by specific phrasing models and then synthesized into speech using TTS models, which were later used in the mean opinion score (MOS) tests. We fixed all random seeds during synthesis to ensure consistent outputs for the same input text.

We conducted MOS tests on Prolific\footnote{\url{https://www.prolific.com/}}. In each test, a native English listener was instructed to pay attention to the effect of inserted pauses on prosody and rhythm and rated the naturalness of each utterance on a five-point scale (1 = bad, 2 = poor, 3 = fair, 4 = good, 5 = excellent). Speech samples used in the MOS tests are available on our demo page\footnote{\url{https://ydqmkkx.github.io/phrasing-demo/}}.

\subsection{Training strategy}
\label{sec: configs training}
All training processes in our experiments were conducted on a single 40~GB Nvidia A100 GPU with a mini-batch size of 32, except for Matcha-TTS in Section~\ref{sec: train tts}, which was trained on 2 $\times$ 96~GB Nvidia H100 GPUs with a mini-batch size of 128. We utilized the AdamW~\cite{loshchilov19adamw} optimizer with PyTorch's default hyper-parameter settings\footnote{\url{https://pytorch.org/docs/stable/generated/torch.optim.AdamW.html}}. We also applied a linear learning rate scheduler in Section~\ref{sec: train phrasing} and Section~\ref{sec: train tts} that warmed up the learning rate over the initial 10\% training steps to a peak, then linearly reduced it to 0.

\subsubsection{Training of phrasing models}
\label{sec: train phrasing}

The phrasing models were trained using the training set described in Section~\ref{sec: dataset} with binary cross-entropy (BCE) loss and single-precision floating-point (FP32) computations. The training process was divided into two stages to ensure training stability: 1) In the first stage, the PLM parameters were frozen, and the model was trained for 10 epochs with a peak learning rate of $5 \times 10^{-4}$. 2) In the second stage, we unfroze the PLM parameters and continue to train the model for 10 epochs with a peak learning rate of $5 \times 10^{-6}$. To enhance stability further, we applied gradient clipping for the PLM with a maximum norm of 1.0. The models were evaluated every 1000 training steps on the validation-seen set, and the weights and corresponding threshold achieving the highest F\textsubscript{0.5} score were saved.

In practice, a single-stage training setup with a linear scheduler and a peak learning rate of $5 \times 10^{-4}$, without freezing any parameters, already yields strong performance for both subword-level BASE PLMs and phoneme-level PLMs. However, to enable fair comparisons with the subword-level LARGE PLMs, which are more susceptible to gradient explosion, we adopted the above two-stage training framework for all models. This design ensures both training stability and efficiency across different PLM sizes and configurations, particularly when training phrasing models with subword-level LARGE PLMs.

\subsubsection{Training of the TTS model}
\label{sec: train tts}
During training and inference, the symbol ``/'' was designated as the RP mark in the text transcripts. For subjective evaluations with seen speakers, only utterances from the ``train-clean-360'' subset in the training and test-seen sets were used to ensure higher synthesis quality. VITS was trained for 400 epochs with a peak learning rate of $2 \times 10^{-4}$, while Matcha-TTS was trained for 800 epochs with a peak learning rate of $4 \times 10^{-4}$. For subjective evaluations with unseen speakers, we used the validation-unseen set to fine-tune VITS for 400 epochs and Matcha-TTS for 800 epochs to generate synthetic speech for these speakers.

\subsubsection{Training of the embedding adapter}
\label{sec: train adapter}

For the proposed phrasing models with trainable PSVM embeddings, we obtained paired sets $\{\mathbf{e}_m\}_{m=0}^{M-1}$ and $\{\mathbf{e}'_m\}_{m=0}^{M-1}$ (as discussed in Section~\ref{sec: proposed adapter}) for seen speakers. These pairs were used as the input and target during training. The adapter was trained for 100,000 steps with a constant learning rate of $1 \times 10^{-5}$. 

\section{Experimental evaluations}

\subsection{Experiments on seen speakers}
\label{sec: seen}
This section explores the impact of different speaker embeddings and PLMs on phrasing performance. The experiments involved seen speakers that were present in the training set. 

\subsubsection{Evaluation of the baseline model}

\begin{table}[t]
  \caption{Phrasing results of the baseline model for seen speakers.}
  \label{tab: baseline seen results}
  \centering
  \small
  \begin{tabular}{c c c}
    \toprule
    Precision & Recall & \textbf{F\textsubscript{0.5}}\\
    \midrule
    0.4309 & 0.2402 & 0.3719\\
    \bottomrule
  \end{tabular}
\end{table}

Phrasing results of the baseline model on the test-seen set are presented in Table~\ref{tab: baseline seen results}. Although the F\textsubscript{0.5} score is the primary metric, corresponding precision and recall values are provided as supplementary information.

\subsubsection{Evaluations on speaker embeddings}
\label{sec: eval spk}

\begin{table*}[t]
  \caption{Phrasing results of the proposed models with various speaker embeddings. EER: equal error rate on Voxceleb 1 clean test set. Voxceleb: Voxceleb 1 and Voxceleb 2. Voxceleb+: Voxceleb with data augmentation. Voxceleb++: Voxceleb, Fisher~\cite{cieri04fisher}, Switchboard~\cite{godfrey93switchboard}, LibriSpeech~\cite{panayotov15librispeech} with data augmentation. \textbf{Bold}: the highest score in each category. \underline{Underlined}: the overall highest score.}

  \label{tab: spk}
  \centering
  \scalebox{0.9}{
  \begin{tabular}{l l c c c c c c c c c}
    \toprule
    \multirow{2}{*}{Speaker embeddings} & \multirow{2}{*}{Training data} & \multirow{2}{*}{Dimension} & \multirow{2}{*}{EER} & \multicolumn{3}{c}{Frozen} & \multicolumn{3}{c}{Trainable}\\ \cmidrule(lr){5-7} \cmidrule(lr){8-10} & & & & Precision & Recall & \textbf{F\textsubscript{0.5}} & Precision & Recall & \textbf{F\textsubscript{0.5}}\\
    \toprule
    Xavier initialization & -- & 192 & -- & -- & -- & -- & 0.5644 & 0.2917 & \underline{\textbf{0.4755}}\\
    ECAPA-TDNN & Voxceleb & 192 & 0.80\% & 0.5212 & 0.2723 & 0.4406 & 0.5276 & 0.2947 & 0.4556\\
    ResNet-TDNN & Voxceleb & 256 & 1.05\% & 0.5322 & 0.2912 & \textbf{0.4566} & 0.5832 & 0.2650 & 0.4702\\
    SpeakerNet & Voxceleb+ & 256 & 1.93\% & 0.5167 & 0.2696 & 0.4367 & 0.5469 & 0.3054 & 0.4722\\
    TitaNet\textsubscript{LARGE} & Voxceleb++ & 192 & 0.68\% & 0.5168 & 0.2438 & 0.4222 & 0.5721 & 0.2828 & 0.4750\\
    \bottomrule
  \end{tabular}
  }
\end{table*}

We experimented with the proposed model using various speaker embeddings. The phrasing results on the test-seen set are presented in Table~\ref{tab: spk}. From them we can draw the following conclusions:
\begin{itemize}
    \item Incorporating speaker embeddings significantly enhanced the F\textsubscript{0.5} score of the proposed model compared to the baseline model. This underscores that different speakers have different habits and styles when inserting RPs. Using speaker embeddings can capture these features effectively, thereby improving phrasing accuracy.
    
    \item Frozen PSVM embeddings positively impacted the results compared to the baseline model, as they captured the prosody and rhythm features of speakers. However, their efficacy did not directly align with their performance in speaker verification tasks. Moreover, trainable embeddings outperformed frozen PSVM embeddings, suggesting that the encoded features in PSVM embeddings could not provide sufficient information for phrasing tasks.
    
    \item When using trainable speaker embeddings, PSVM embeddings did not improve phrasing accuracy compared to randomly initialized embeddings. This suggests that trainable, randomly initialized speaker embeddings are both convenient and effective for phrasing tasks involving seen speakers.
\end{itemize}

\subsubsection{Evaluations on PLMs}
\label{sec: eval plm}

\begin{table*}[t]
  \caption{Phrasing results of the proposed models with various PLMs. Model size: the number of parameters in millions (M).}
  \label{tab: plms}
  \centering
  \small
  \begin{tabular}{l r r c c c c c c c}
    \toprule
    PLM & Hidden size & Model size & Precision & Recall & \textbf{F\textsubscript{0.5}}\\
    \toprule
    BERT\textsubscript{BASE} & 768 & 110M & 0.5644 & 0.2917 & 0.4755\\
    XLNET\textsubscript{BASE} & 768 & 110M & 0.5624 & 0.2826 & 0.4695\\
    RoBERTa\textsubscript{BASE} & 768 & 125M & 0.5691 & 0.2906 & 0.4776\\
    ALBERT\textsubscript{BASE} & 768 & 11M & 0.5476 & 0.2857 & 0.4627\\
    DeBERTaV3\textsubscript{BASE} & 768 & 184M & 0.5619 & 0.2992 & \textbf{0.4780}\\
    \midrule
    BERT\textsubscript{LARGE} & 1,024 & 336M & 0.5854 & 0.2737 & 0.4768\\
    XLNET\textsubscript{LARGE} & 1,024 & 340M & 0.5808 & 0.2742 & 0.4746\\
    RoBERTa\textsubscript{LARGE} & 1,024 & 355M & 0.5801 & 0.2957 & \textbf{0.4865}\\
    ALBERT\textsubscript{LARGE} & 1,024 & 17M & 0.5714 & 0.2763 & 0.4708\\
    DeBERTaV3\textsubscript{LARGE} & 1,024 & 434M & 0.5769 & 0.2781 & 0.4748\\
    \midrule
    MP BERT & 768 & 111M & 0.6238 & 0.2773 & \underline{\textbf{0.4991}}\\
    PL BERT & 768 & 88M & 0.5826 & 0.2919 & 0.4858\\
    \bottomrule
  \end{tabular}
\end{table*}

The best-performing phrasing model in Section~\ref{sec: eval spk}, which incorporates randomly initialized speaker embeddings, was used as the backbone of this section. We employed various PLMs to investigate the potential enhancements they could offer to predictive accuracy. The experimental results presented in Table~\ref{tab: plms} reveal some insights:

\begin{itemize}
    \item Among the subword-level PLMs with base sizes, DeBERTaV3\textsubscript{BASE} achieved the highest performance. For the subword-level PLMs with large sizes, RoBERTa\textsubscript{LARGE} performed the best, with an F\textsubscript{0.5} score of 0.4865. These more advanced subword-level PLMs provided richer and more powerful linguistic features and improved phrasing accuracy.
    
    \item Except for ALBERT with a unique architecture, the large variants of these subword-level PLMs contain approximately three times as many parameters as their base variants. Despite the substantial increase in complexity, the improvements in phrasing accuracy are relatively modest. This suggests that the exploitation of subword-level features may have reached its limits within the current framework of the phrasing model.
    
    \item When using phoneme-level PLMs, PL BERT achieved a similar F\textsubscript{0.5} score to RoBERTa\textsubscript{LARGE}, while MP BERT outperformed all other PLMs with an F\textsubscript{0.5} score of 0.4991. These results indicate that phoneme-level PLMs hold certain advantages over subword-level PLMs for this task. The sup-phoneme method utilized in MP BERT effectively enhances representation capabilities, leading to better performance than PL BERT.
    
    \item Due to computing resource constraints, our implementations used BookCorpus~\cite{zhu15bookcorpus} plus English Wikipedia to pre-train the two phoneme-level PLMs for 10 epochs, which is fewer than other subword-level PLMs. Despite this limitation, their outstanding performance demonstrates that applying phoneme-level PLMs to TTS front-end tasks and other text-processing tasks is promising and merits further exploration.
\end{itemize}

\subsubsection{Subjective evaluations}
\label{sec: subjective seen}

\begin{table*}[t]
  \caption{MOS for seen speakers. CI: 95\% confidence interval.}
  \label{tab: mos seen}
  \centering
  \small
  \begin{tabular}{ l l l l c c c}
    \toprule
    \multicolumn{4}{c}{Phrasing method} & \multirow{3}{*}{F\textsubscript{0.5}$\uparrow$} & \multicolumn{2}{c}{MOS$\uparrow$ $\pm$ CI}\\
    \cmidrule(lr){1-4} \cmidrule(lr){6-7} \multirow{2}{*}{Model} & \multirow{2}{*}{PLM} & \multicolumn{2}{c}{Speaker embeddings} & & \multirow{2}{*}{VITS} & \multirow{2}{*}{Matcha-TTS}\\
    \cmidrule(lr){3-4} & & Initialization & Parameters\\
    \toprule
    Ground-truth RP & -- & -- & -- & -- & 3.03 $\pm$ 0.16 & 3.48 $\pm$ 0.15\\
    \hdashline
    \addlinespace[0.5ex]
    Baseline & BERT\textsubscript{BASE} & -- & -- & 0.3719 & 2.74 $\pm$ 0.15 & 3.35 $\pm$ 0.16\\
    Proposed & BERT\textsubscript{BASE} & ResNet-TDNN & Frozen & 0.4566 & 2.88 $\pm$ 0.15 & 3.41 $\pm$ 0.15\\
    Proposed & BERT\textsubscript{BASE} & Xavier & Trainable & 0.4755 & 2.91 $\pm$ 0.16 & 3.42 $\pm$ 0.15\\
    Proposed & MP BERT & Xavier & Trainable & \textbf{0.4991} & \textbf{2.98 $\pm$ 0.16} & \textbf{3.51 $\pm$ 0.14}\\
    \bottomrule
  \end{tabular}
\end{table*}

For the MOS tests, we randomly selected 100 utterances from the test-seen set, each containing between 30 and 60 words. As displayed in Table~\ref{tab: mos seen}, their text transcripts were phrased using the five phrasing methods to insert RP marks and yield 500 speech utterances:

\begin{itemize}
    \item Ground-truth RP: Instead of using ground-truth speech utterances directly, we synthesized the utterances with the ground-truth RP marks to eliminate the quality gap between synthetic and natural speech.
    
    \item Baseline model.
    
    \item Best-performing proposed models in Section~\ref{sec: eval spk}: One model uses frozen speaker embeddings, and the other uses trainable speaker embeddings.
    
    \item Best-performing proposed model in Section~\ref{sec: eval plm}: This model uses MP BERT as the PLM module.
\end{itemize}

Each test contained 20 speech samples. For synthetic speech generated by VITS, we distributed 50 tests and obtained 48 valid responses. For synthetic speech from Matcha-TTS, we also distributed 50 tests and received 50 valid responses. The results are summarized in Table~\ref{tab: mos seen}. The MOS for these phrasing methods generally correlates with their F\textsubscript{0.5} score on the test set. According to the t-test analysis, only the proposed model utilizing trainable speaker embeddings and MP BERT outperformed the baseline with statistical significance, while only the baseline significantly underperformed the ground truth.

\subsection{Experiments on unseen speakers}
\label{sec: unseen}

In previous sections, we pre-trained the phrasing models with a large amount of data in the training set and focused on the phrasing task for seen speakers. However, our proposed model is not directly applicable to unseen speakers, for which we prefer a solution that employs pre-trained phrasing models and avoids fine-tuning due to the intensive computational cost. With this method, our options for addressing unseen speakers are limited to either the baseline or the proposed models with PSVM embeddings, the latter offering the potential for enhancing phrasing accuracy through few-shot learning. Therefore, this section concentrates on the unseen speakers and proposes a few-shot adaptation method without fine-tuning. 

Due to the success of MP BERT discussed in Section~\ref{sec: eval plm}, we also pre-trained the baseline model and proposed models with MP BERT (\ref{appendix e}) and utilized them in this section. In the experimental results, we only present the primary metric, F\textsubscript{0.5} score, to ensure clarity.

\subsubsection{Evaluation of the baseline model}

\begin{table}[t]
  \caption{Phrasing results of the baseline model for unseen speakers.}
  \label{tab: baseline unseen results}
  \centering
  \small
  \begin{tabular}{l c}
    \toprule
    PLM & F\textsubscript{0.5}\\
    \midrule
    BERT\textsubscript{BASE} & \textbf{0.3188}\\
    MP BERT & 0.3069\\
    \bottomrule
  \end{tabular}
\end{table}

Since the baseline model does not incorporate speaker embeddings, it can be directly applied to the test-unseen set after pre-training. The phrasing results are presented in Table~\ref{tab: baseline seen results}, where we experimented with both BERT\textsubscript{BASE} and MP BERT for the baseline model:
\begin{itemize}
    \item Both models showed a significant decrease in the F\textsubscript{0.5} score on the test-unseen set compared to the test-seen set, indicating a clear difference in speaker distribution between the test-unseen set and the training/test-seen sets for the phrasing task.

    \item For the baseline model, BERT\textsubscript{BASE} outperformed MP BERT on the test-unseen set. Based on the results in Table~\ref{tab: baseline unseen results} and Table~\ref{tab: mpbert} (a), MP BERT's optimized performance on seen speakers may limit its generalization ability for unseen speakers, whose RP distribution is different from that of seen speakers (\ref{appendix f}).
\end{itemize}

\subsubsection{Evaluation on few-shot adaptation}
\label{sec: eval few-shot}

\begin{table*}[t]
  \caption{Phrasing results of few-shot adaptation for unseen speakers.}
  \label{tab: few-shot}
  \centering
  \small

  \begin{subtable}{1.0\linewidth}
    \caption{\footnotesize Proposed models with frozen speaker embeddings.}
    \centering
    \small
      \begin{tabular}{l l c c c c c c c c}
        \toprule
        \multirow{2}{*}{PLM} & \multirow{2}{*}{Speaker embedding} &\multicolumn{7}{c}{Samples} \\ \cmidrule(lr){3-9} & & 1 & 5 & 10 & 20 & 30 & 40 & 50\\
        \toprule
        \multirow{4}{*}{BERT\textsubscript{BASE}}
        & ECAPA-TDNN & 0.2264 & 0.3074 & 0.3255 & 0.3355 & 0.3480 & 0.3559 & 0.3589\\
        & ResNet-TDNN & 0.2497 & 0.3277 & 0.3357 & 0.3419 & 0.3584 & 0.3585 & 0.3602\\
        & SpeakerNet & 0.2665 & 0.3326 & 0.3553 & \textbf{0.3670} & 0.3760 & \textbf{0.3761} & \textbf{0.3728}\\
        & TitaNet\textsubscript{LARGE} & \textbf{0.2845} & \textbf{0.3491} & \textbf{0.3583} & 0.3656 & \underline{\textbf{0.3784}} & 0.3724 & 0.3708\\
        \midrule
        \multirow{4}{*}{MP BERT} 
        & ECAPA-TDNN & 0.2321 & 0.3334 & 0.3387 & 0.3515 & \textbf{0.3651} & \underline{\textbf{0.3758}} & \textbf{0.3712}\\
        & ResNet-TDNN & 0.2661 & 0.3418 & 0.3479 & 0.3463 & 0.3617 & 0.3603 & 0.3598\\
        & SpeakerNet & 0.2664 & 0.3155 & 0.3307 & 0.3447 & 0.3568 & 0.3577 & 0.3600\\
        & TitaNet\textsubscript{LARGE} & \textbf{0.2815} & \textbf{0.3485} & \textbf{0.3530} & \textbf{0.3522} & 0.3626 & 0.3556 & 0.3590\\
        \bottomrule
      \end{tabular}
    \end{subtable}
    \vspace{5pt} 
    
  \begin{subtable}{1.0\linewidth}
    \caption{\footnotesize Proposed models with trainable speaker embeddings.}
    \centering
    \small
      \begin{tabular}{l l c c c c c c c c}
        \toprule
        \multirow{2}{*}{PLM} & \multirow{2}{*}{Speaker embedding} &\multicolumn{7}{c}{Samples} \\ \cmidrule(lr){3-9} & & 1 & 5 & 10 & 20 & 30 & 40 & 50\\
        \toprule
        \multirow{4}{*}{BERT\textsubscript{BASE}}
        & ECAPA-TDNN & 0.2406 & 0.3120 & 0.3232 & 0.3499 & 0.3633 & 0.3630 & 0.3695\\
        & ResNet-TDNN & 0.2366 & 0.3319 & \textbf{0.3718} & \textbf{0.3845} & \textbf{0.4011} & \underline{\textbf{0.4041}} & \textbf{0.4007}\\
        & SpeakerNet & 0.2852 & 0.3253 & 0.3379 & 0.3545 & 0.3588 & 0.3608 & 0.3611\\
        & TitaNet\textsubscript{LARGE} & \textbf{0.2980} & \textbf{0.3591} & 0.3617 & 0.3595 & 0.3741 & 0.3709 & 0.3711\\
        \midrule
        \multirow{4}{*}{MP BERT} 
        & ECAPA-TDNN & 0.2865 & 0.3182 & 0.3176 & 0.3437 & 0.3548 & 0.3551 & 0.3534\\
        & ResNet-TDNN & 0.2511 & 0.3381 & \textbf{0.3546} & \textbf{0.3566} & \underline{\textbf{0.3723}} & \textbf{0.3707} & \textbf{0.3719}\\
        & SpeakerNet & \textbf{0.3111} & 0.3382 & 0.3515 & 0.3565 & 0.3652 & 0.3676 & 0.3693\\
        & TitaNet\textsubscript{LARGE} & 0.2983 & \textbf{0.3518} & 0.3544 & 0.3555 & 0.3664 & 0.3662 & 0.3666\\
        \bottomrule
      \end{tabular}
    \end{subtable}
\end{table*}

To investigate the few-shot adaptation approach, we randomly sampled utterances from each speaker in the validation-unseen set at sizes of 1, 5, 10, 20, 30, 40, and 50. For each sample size, we extracted and averaged each speaker's embeddings using various PSVMs, which replaced the speaker embedding layer in the proposed models with frozen embeddings. For models using trainable speaker embeddings, the PSVM embeddings were adapted through the embedding adapter described in Section~\ref{sec: proposed adapter}. During evaluation, the same thresholds obtained from pre-training on the validation-seen set were applied. The phrasing results on the test-unseen set are shown in Table~\ref{tab: few-shot}, which reveals the following insights:
\begin{itemize}
    \item When using only one utterance per speaker, the proposed models with few-shot adaptation performed weaker than the baseline model. However, with an increase to five utterances, the proposed models generally surpassed the baseline models, except for the model using ECAPA-TDNN embeddings that required ten utterances. This validates the effectiveness of few-shot adaptation in this phrasing task.
    
    \item As the sample count increased, the performance of the proposed models generally improved, suggesting that more samples provided more representative information and enhanced predictive accuracy. However, beyond a certain point, further increases in sample size did not lead to additional improvements and even caused performance degradation due to excessive noise.

    \item With few-shot adaptation, proposed models using BERT\textsubscript{BASE} generally outperformed their counterparts using MP BERT in terms of F\textsubscript{0.5} score. This could be attributed to the limited generalization capabilities of MP BERT, as similar observed in the baseline models.
\end{itemize}

\subsubsection{Subjective evaluations}

\begin{table*}[t]
  \caption{MOS for unseen speakers. CI: 95\% confidence interval.}
  \label{tab: mos unseen}
  \centering
  \small
  \begin{tabular}{ l l l l c c c c}
    \toprule
    \multicolumn{5}{c}{Phrasing method} & \multirow{3}{*}{F\textsubscript{0.5}$\uparrow$} & \multicolumn{2}{c}{MOS$\uparrow$ $\pm$ CI}\\
    \cmidrule(lr){1-5} \cmidrule(lr){7-8}\multirow{2}{*}{Model} & \multirow{2}{*}{PLM} & \multicolumn{2}{c}{Speaker embedding} & \multirow{2}{*}{Samples} & & \multirow{2}{*}{VITS} & \multirow{2}{*}{Matcha-TTS}\\
    \cmidrule(lr){3-4} & & Initialization & Parameters\\
    \toprule
    Ground-truth RP & -- & -- & -- & 0 & -- & 3.22 $\pm$ 0.15 & 3.58 $\pm$ 0.14 \\
    \hdashline
    \addlinespace[0.5ex]
    Baseline & BERT\textsubscript{BASE} & -- & -- & 0 & 0.3188 & 3.02 $\pm$ 0.17 & 3.39 $\pm$ 0.15 \\
    Baseline & MP BERT & -- & -- & 0 & 0.3069 & 2.99 $\pm$ 0.16 & 3.30 $\pm$ 0.15 \\
    Proposed & BERT\textsubscript{BASE} & Titanet\textsubscript{LARGE} & Frozen & 30 & 0.3784 & 3.08 $\pm$ 0.15 & 3.42  $\pm$ 0.14 \\
    Proposed & MP BERT & ECAPA-TDNN & Frozen & 40 & 0.3758 & 3.11 $\pm$ 0.15 & 3.46 $\pm$ 0.14 \\
    Proposed & BERT\textsubscript{BASE} & ResNet-TDNN & Trainable & 40 & \textbf{0.4041} & 3.14 $\pm$ 0.16 & \textbf{3.47 $\pm$ 0.14} \\
    Proposed & MP BERT & ResNet-TDNN & Trainable & 30 & 0.3723 & \textbf{3.19 $\pm$ 0.17} & \textbf{3.47 $\pm$ 0.14} \\
    \bottomrule
  \end{tabular}
\end{table*}

In the MOS tests, we evaluated the impact of the following phrasing methods: the ground-truth RP, the baseline models, and four proposed models with few-shot adaptation that were underlined in Table~\ref{tab: few-shot}. Considering the number of evaluated phrasing methods, each test included 28 sampled utterances. For synthetic speech generated by VITS and Matcha-TTS, we distributed 50 tests and obtained 50 valid results for both, which are summarized in Table~\ref{tab: mos unseen}:
\begin{itemize}
    \item The proposed models with few-shot adaptation outperformed the baseline model. Besides, the proposed models using trainable speaker embeddings and adapters received higher MOS than those with frozen speaker embeddings.

    \item Interestingly, with few-shot adaptation, the MOS for the proposed models using MP BERT was not directly proportional to their F\textsubscript{0.5} score. These models achieved similar or higher MOS than their counterparts using BERT\textsubscript{BASE}. This discrepancy indicates that applying MP BERT to our proposed model may improve the alignment of inserted RPs with human preferences.

    \item According to the t-test analysis, only the baseline models significantly underperformed the ground truth with statistical significance.
\end{itemize}

\section{Analysis and discussions}
\label{sec: discussion}

In this section, we first present a graphical analysis of the RP distribution across speakers in the training set (Section~\ref{sec: rp distribution}). Then, in Section~\ref{sec: speaker analysis}, we investigate the speaker characteristics captured by different types of speaker embeddings.

\subsection{Overall RP distribution across speakers}
\label{sec: rp distribution}

\begin{figure*}
    \centering
    \includegraphics[width=1.0\linewidth, clip]{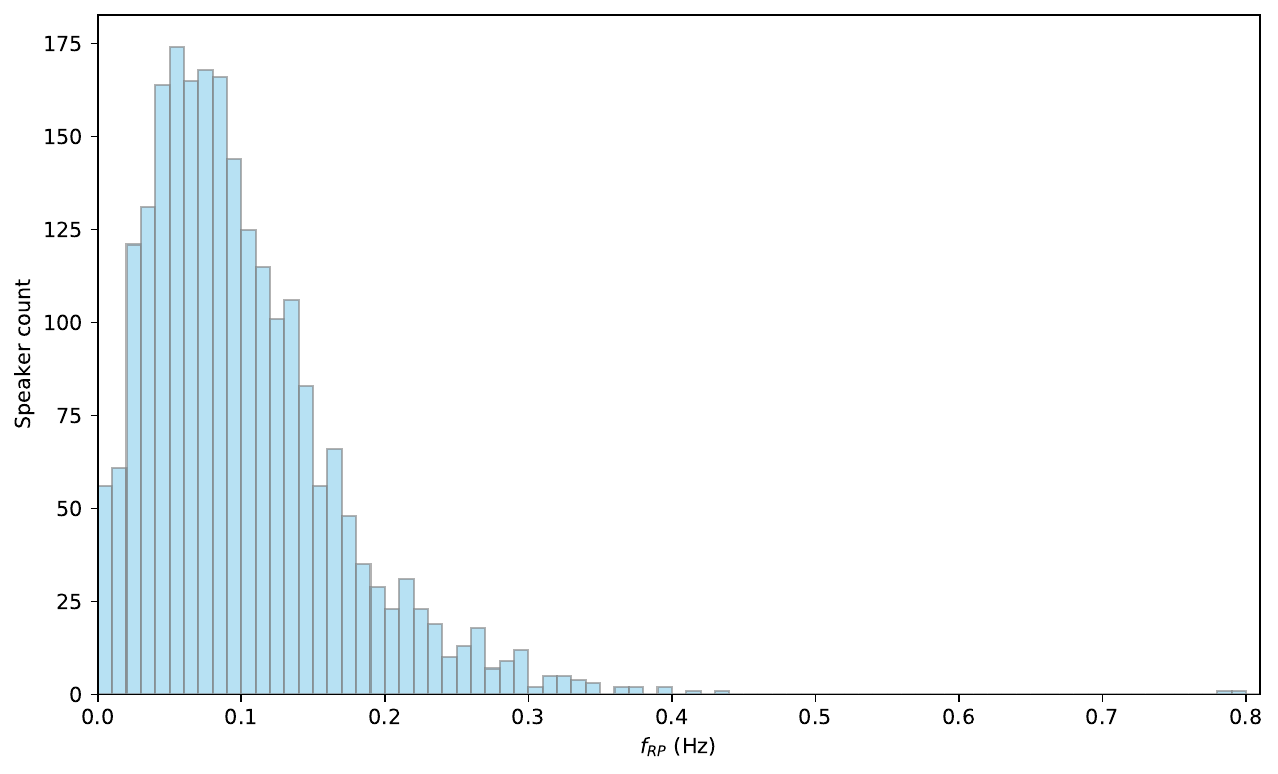}
    \caption{Distribution of RP insertion frequency across speakers.}
    \label{fig: rp frequency}
\end{figure*}

\begin{figure*}
    \centering
    \includegraphics[width=1.0\linewidth, clip]{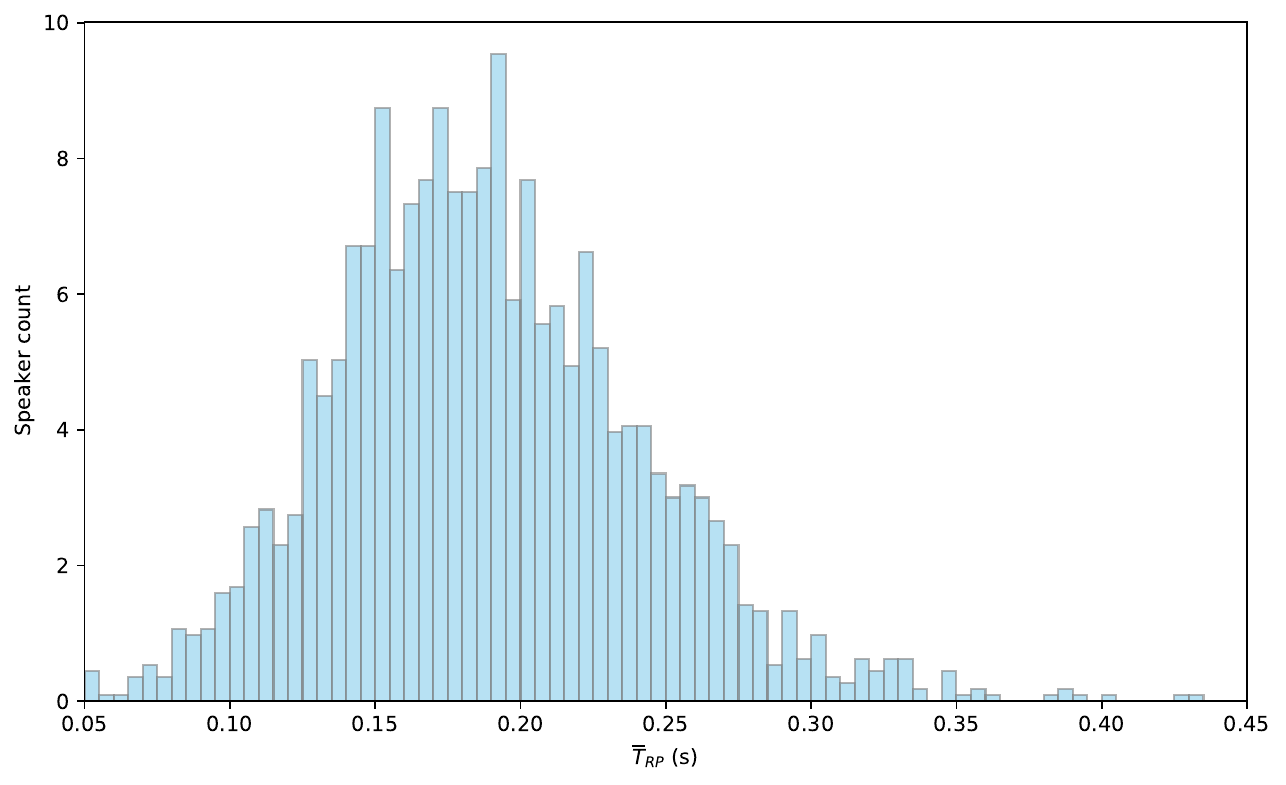}
    \caption{Distribution of average RP duration across speakers.}
    \label{fig: rp duration}
\end{figure*}

We present the overall RP distribution across speakers in the training set to provide a direct and intuitive understanding of their differences in phrasing. For each speaker, we calculate the frequency $f_{RP}$ (in Hz) and average duration $\overline{T}_{RP}$ (in seconds) of RP insertion:

\begin{equation}
    f_{RP} = \frac{N_{RP}}{T_{utterance}}, 
    \overline{T}_{RP} = \frac{T_{RP}}{N_{RP}},
\end{equation}

\noindent where $N_{RP}$ denotes the total number of RPs, $T_{RP}$ represents the total duration of RPs, and $T_{utterance}$ is the total duration of all utterances for the speaker. The distributions of $f_{RP}$ and $\overline{T}_{RP}$ across speakers are illustrated in Fig.~\ref{fig: rp frequency} and Fig.~\ref{fig: rp duration}, respectively. In the latter, speakers without any RP occurrences are excluded.

\subsection{Speaker embedding analysis in phrasing models}
\label{sec: speaker analysis}

For phrasing models with a randomly initialized speaker embedding layer, the speaker embeddings (referred to as phrasing speaker embeddings) are only exposed to textual inputs and RP positions during training. In this subsection, we investigate the latent space distribution of phrasing speaker embeddings and PSVM embeddings in relation to speaker characteristics. It is important to note that due to the different training objectives, phrasing speaker embeddings primarily serve to distinguish between speakers (functioning similarly to one-hot vectors) and encode limited speaker-specific information, while other components of the phrasing model also store some speaker-relevant features~\cite{kocmi17randomembed}. Therefore, phrasing speaker embeddings are less effective than PSVM embeddings in representing speaker characteristics. Here, we attempt to capture these relatively weak representations and explore which speaker characteristics the phrasing models can infer from plain text alone.

We extracted the speaker embedding layers from the bottom three phrasing models in Table~\ref{tab: mos seen} and obtained the corresponding speaker embeddings -- specifically, the ResNet-TDNN embeddings and two types of phrasing speaker embeddings. Then, we applied $k$-means clustering~\cite{kmeans} to each embedding set separately.

\subsubsection{Visualization}

\begin{figure*}
    \centering
    \includegraphics[width=0.95\linewidth, clip]{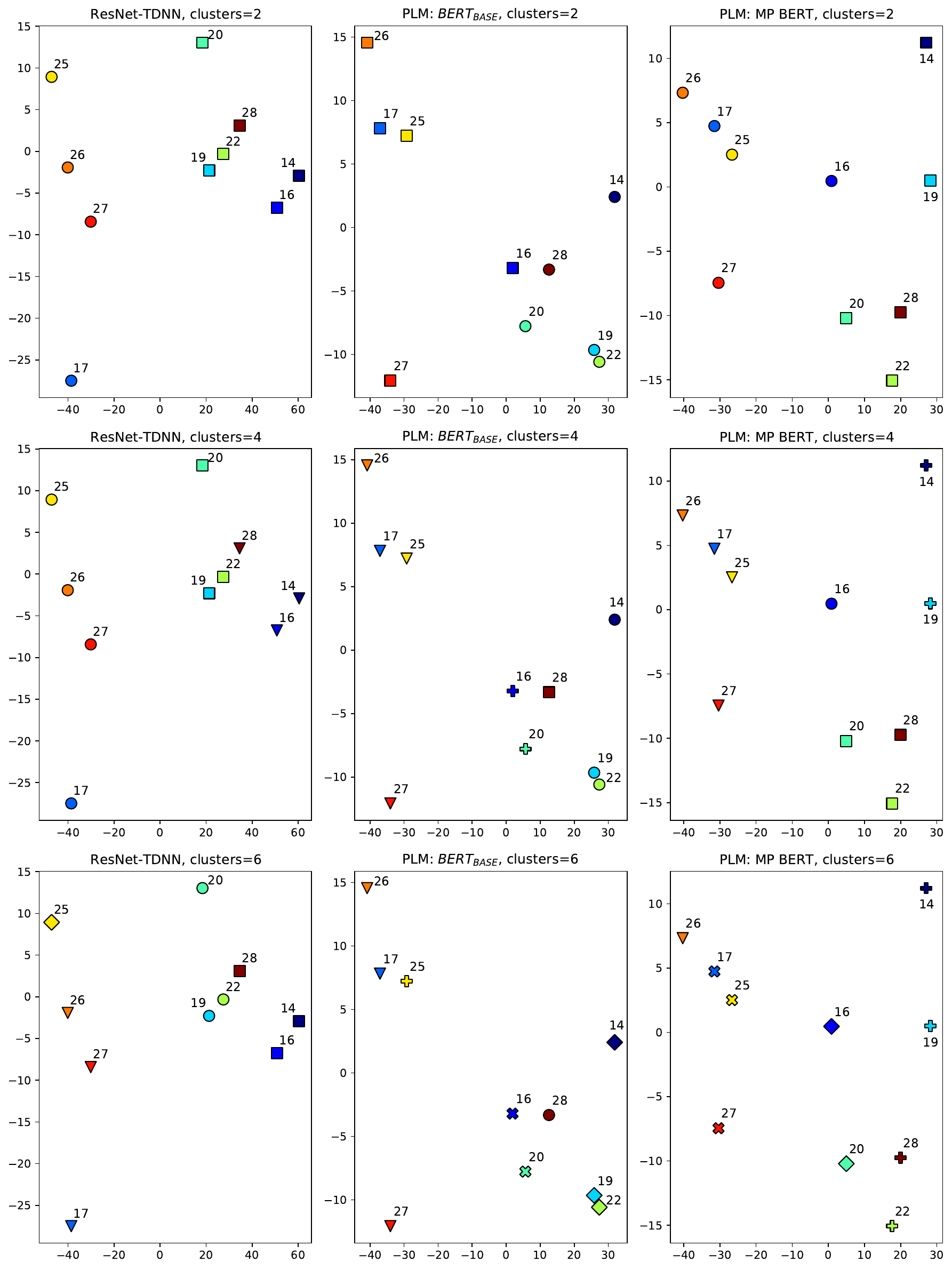}
    \caption{Visualization of speaker embeddings with t-SNE.}
    \label{fig: tsne}
\end{figure*}

For visualization, we set the number of clusters $k$ to relatively small values: 2, 4, and 6. Then, we employed t-SNE~\cite{tsne} to project the speaker embeddings into a two-dimensional space. As depicted in Fig.~\ref{fig: tsne}, we select the first 10 speakers by ID order for demonstration. Additionally, we provide human-annotated speaker characteristics in Table~\ref{tab: speaker prompt}, which are derived from LibriTTS-P~\cite{libritts-p}\footnote{Annotator 1: \url{https://github.com/line/LibriTTS-P/blob/main/data/df1_en.csv}}. We can observe that:

\begin{itemize}
    \item When $k=2$, $k$-means clustering appeared to reflect gender expression in all three types of embeddings. The ResNet-TDNN embeddings showed a strict separation aligned with gender categories, while the two phrasing speaker embeddings exhibited a similar trend, except for speaker 16. This suggests a potential correlation between RP insertion and gender expression.

    \item When $k=4$, the clusters reflected gender-based separation more clearly, with speakers expressing femininity further divided. Besides, the clustering results began to diverge across different types of embeddings.

    \item When $k=6$, speaker 19 and speaker 22 were consistently clustered together across all three embeddings. They are described as \textit{powerful}, \textit{fluent}, \textit{(slightly) raspy}, \textit{intellectual}, \textit{(slightly) calm}, \textit{friendly}, \textit{reassuring}, \textit{elegant}, and \textit{(slightly) kind}. Similarly, speaker 17 and speaker 27 were grouped into the same cluster and shared annotations such as \textit{cool}, \textit{calm}, and \textit{wild}.
\end{itemize}

This visualization reveals that, despite the training objective of phrasing models differing from PSVMs, the extracted phrasing speaker embeddings still exhibit certain correlations with annotated speaker characteristics. While clustering results show both commonalities and differences among the three types of embeddings, these patterns warrant further investigation, which we pursue in the next part.

\begin{table*}[t]
  \caption{Prompts of speaker characteristics from LibriTTS-P.}
  \label{tab: speaker prompt}
  \centering
  \small
  \begin{tabular}{r >{\raggedright\arraybackslash}p{14cm}}
    \toprule
    ID & Prompts \\
    \midrule
    14 & feminine, slightly gender-neutral, young, tensed, powerful, bright, clear, fluent, cute, friendly, unique, lively, slightly \\
    16 & feminine, young, tensed, bright, slightly soft, clear, fluent, slightly intellectual, friendly, lively, slightly kind \\
    17 & masculine, middle-aged, thick, powerful, slightly dark, muffled, raspy, cool, intellectual, calm, mature, wild, slightly strict, slightly sharp \\
    19 & feminine, middle-aged, tensed, powerful, clear, fluent, slightly raspy, intellectual, calm, friendly, reassuring, elegant, lively, slightly kind \\
    20 & feminine, middle-aged, relaxed, slightly weak, slightly dark, muffled, slightly raspy, slightly intellectual, calm, slightly kind, modest \\
    22 & feminine, adult-like, slightly relaxed, powerful, slightly soft, fluent, raspy, intellectual, slightly calm, friendly, reassuring, elegant, kind, modest \\
    25 & masculine, adult-like, tensed, slightly powerful, bright, clear, intellectual, friendly, reassuring, lively, kind \\
    26 & masculine, adult-like, thick, tensed, powerful, slightly hard, clear, fluent, calm, slightly mature, friendly, reassuring, wild, lively, slightly kind \\
    27 & masculine, adult-like, tensed, slightly bright, clear, fluent, cool, calm, wild, lively \\
    28 & feminine, middle-aged, relaxed, slightly dark, clear, slightly nasal, intellectual, calm, slightly friendly, reassuring, elegant, slightly sharp, modest \\
    \bottomrule
  \end{tabular}
\end{table*}

\begin{table}
  \caption{Results of chi-squared tests (ResNet-TDNN).}
  \label{tab: chi resnet}
  \centering
  \small
  \begin{tabular}{l r l r r l r r l r}
    \toprule
    \multirow{2}{*}{Characteristic} & \multicolumn{3}{c}{$k=4$} & \multicolumn{3}{c}{$k=16$} & \multicolumn{3}{c}{$k=64$}\\
    \cmidrule(lr){2-4} \cmidrule(lr){5-7} \cmidrule(lr){8-10} & $\chi^2$ & $p$ & $V$ & $\chi^2$ & $p$ & $V$ & $\chi^2$ & $p$ & $V$ \\
    \midrule
    Gender expression & 2223.92 & $<$ 0.0001 & 0.57 
                      & 2323.99 & $<$ 0.0001 & 0.58
                      & 2419.72 & $<$ 0.0001 & 0.59 \\
    Age            & 601.28 & $<$ 0.0001 & 0.30
                   & 1039.83 & $<$ 0.0001  & 0.34 
                   & 1343.49 & $<$ 0.0001 & 0.38 \\
    \hdashline
    \addlinespace[0.5ex]
    wild            & 1456.27  & $<$ 0.0001  & 0.80 
                    & 1510.56  & $<$ 0.0001  & 0.81 
                    & 1546.89  & $<$ 0.0001  & 0.82 \\
    elegant         & 1030.04  & $<$ 0.0001  & 0.67 
                    & 1128.58  & $<$ 0.0001  & 0.70 
                    & 1202.66  & $<$ 0.0001  & 0.72 \\
    cool            & 719.13   & $<$ 0.0001  & 0.56 
                    & 791.46   & $<$ 0.0001  & 0.59 
                    & 823.35   & $<$ 0.0001  & 0.60 \\
    mature          & 630.79   & $<$ 0.0001  & 0.52 
                    & 803.42   & $<$ 0.0001  & 0.59 
                    & 907.38   & $<$ 0.0001  & 0.63 \\
    sharp           & 370.04   & $<$ 0.0001  & 0.40 
                    & 416.15   & $<$ 0.0001  & 0.43 
                    & 519.89   & $<$ 0.0001  & 0.48 \\
    cute            & 337.97   & $<$ 0.0001  & 0.38 
                    & 597.75   & $<$ 0.0001  & 0.51 
                    & 690.13   & $<$ 0.0001  & 0.55 \\
    sincere         & 221.56   & $<$ 0.0001  & 0.31 
                    & 230.58   & $<$ 0.0001  & 0.32 
                    & 280.74   & $<$ 0.0001  & 0.35 \\
    thick           & 206.76   & $<$ 0.0001  & 0.30 
                    & 378.08   & $<$ 0.0001  & 0.41 
                    & 461.59   & $<$ 0.0001  & 0.45 \\
    fluent          & 166.80   & $<$ 0.0001  & 0.27
                    & 121.22   & $<$ 0.0001  & 0.23 
                    & 189.80   & $<$ 0.0001  & 0.29 \\
    soft            & 131.42   & $<$ 0.0001  & 0.24 
                    & 198.87   & $<$ 0.0001  & 0.29 
                    & 343.61   & $<$ 0.0001  & 0.39 \\
    sweet           & 129.90   & $<$ 0.0001  & 0.24 
                    & 215.93   & $<$ 0.0001  & 0.31 
                    & 271.84   & $<$ 0.0001  & 0.34 \\
    intellectual    & 129.64   & $<$ 0.0001  & 0.24 
                    & 196.50   & $<$ 0.0001  & 0.29 
                    & 255.17   & $<$ 0.0001  & 0.33 \\
    tensed          & 87.17    & $<$ 0.05    & 0.19 
                    & 111.27   & $<$ 0.0001  & 0.22 
                    & 251.40   & $<$ 0.0001  & 0.33 \\
    bright          & 74.78    & $<$ 0.0001  & 0.18 
                    & 201.47   & $<$ 0.0001  & 0.30 
                    & 264.50   & $<$ 0.0001  & 0.34 \\
    muffled         & 63.95    & $<$ 0.0001  & 0.17 
                    & 105.55   & $<$ 0.0001  & 0.21 
                    & 188.67   & $<$ 0.0001  & 0.29 \\
    strict          & 61.48    & $<$ 0.0001  & 0.16 
                    & 157.36   & $<$ 0.0001  & 0.26 
                    & 157.36   & $<$ 0.0001  & 0.26 \\
    friendly        & 52.49    & $<$ 0.0001  & 0.15 
                    & 117.06   & $<$ 0.0001  & 0.23 
                    & 209.17   & $<$ 0.0001  & 0.30 \\
    clear           & 52.23    & $<$ 0.0001  & 0.15 
                    & 134.56   & $<$ 0.0001  & 0.24 
                    & 276.88   & $<$ 0.0001  & 0.35 \\
    refreshing      & 49.54    & $<$ 0.0001  & 0.15 
                    & 132.76   & $<$ 0.0001  & 0.24 
                    & 215.22   & $<$ 0.0001  & 0.31 \\
    hard            & 47.37    & $<$ 0.0001  & 0.14 
                    & 123.94   & $<$ 0.0001  & 0.23 
                    & 182.88   & $<$ 0.0001  & 0.28 \\
    kind            & 45.65    & $<$ 0.0001  & 0.14 
                    & 96.18    & $<$ 0.0001  & 0.20 
                    & 229.48   & $<$ 0.0001  & 0.32 \\
    raspy           & 43.92    & $<$ 0.0001  & 0.14 
                    & 113.74   & $<$ 0.0001  & 0.22 
                    & 253.08   & $<$ 0.0001  & 0.33 \\
    halting         & 16.82    & $<$ 0.005   & 0.09 
                    & 30.40    & $<$ 0.05    & 0.12 
                    & 86.66    & $<$ 0.05    & 0.19 \\
    dark            & 16.40    & $<$ 0.005   & 0.08 
                    & 53.42    & $<$ 0.0001  & 0.15 
                    & 201.59   & $<$ 0.0001  & 0.30 \\
    relaxed         & 14.14    & $<$ 0.005   & 0.08 
                    & 130.93   & $<$ 0.0001  & 0.24 
                    & 287.62   & $<$ 0.0001  & 0.35 \\
    intense         & 12.37    & $<$ 0.01    & 0.07 
                    & 57.63    & $<$ 0.0001  & 0.16 
                    & 135.37   & $<$ 0.0001  & 0.24 \\
    modest          & 11.96    & $<$ 0.01    & 0.07 
                    & 103.09   & $<$ 0.0001  & 0.21 
                    & 280.94   & $<$ 0.0001  & 0.35 \\
    powerful        & 7.91     & $<$ 0.05    & 0.06 
                    & 61.08    & $<$ 0.0001  & 0.16 
                    & 170.21   & $<$ 0.0001  & 0.27 \\
    weak            & --       & --          & --
                    & 72.16    & $<$ 0.0001  & 0.18 
                    & 210.26   & $<$ 0.0001  & 0.30 \\
    calm            & --       & --          & --
                    & 59.25    & $<$ 0.0001  & 0.16 
                    & 162.34   & $<$ 0.0001  & 0.27 \\
    lively          & --       & --          & --
                    & 58.69    & $<$ 0.0001  & 0.16 
                    & --       & --          & --   \\
    unique          & --       & --          & --
                    & 51.43    & $<$ 0.0001  & 0.15 
                    & 140.07   & $<$ 0.0001  & 0.25 \\
    nasal           & --       & --          & --
                    & 35.39    & $<$ 0.005   & 0.12 
                    & 104.89   & $<$ 0.005   & 0.21 \\
    reassuring      & --       & --          & --
                    & 33.22    & $<$ 0.005   & 0.12 
                    & 111.96   & $<$ 0.0005  & 0.22 \\
    thin            & --       & --          & --
                    & 28.78    & $<$ 0.05    & 0.11 
                    & --       & --          & --   \\
    sexy            & --       & --          & --
                    & --       & --          & --
                    & 609.41   & $<$ 0.0001  & 0.51 \\
    light           & --       & --          & --
                    & --       & --          & --
                    & 84.27    & $<$ 0.05    & 0.19 \\
    \bottomrule
  \end{tabular}
\end{table}

\begin{table}
  \caption{Results of chi-squared tests (PLM: BERT\textsubscript{BASE}).}
  \label{tab: chi bert}
  \centering
  \small
  \begin{tabular}{l r l r r l r r l r}
    \toprule
    \multirow{2}{*}{Characteristic} & \multicolumn{3}{c}{$k=4$} & \multicolumn{3}{c}{$k=16$} & \multicolumn{3}{c}{$k=64$}\\
    \cmidrule(lr){2-4} \cmidrule(lr){5-7} \cmidrule(lr){8-10} & $\chi^2$ & $p$ & $V$ & $\chi^2$ & $p$ & $V$ & $\chi^2$ & $p$ & $V$ \\
    \midrule
    Gender expression & 29.91 & $<$ 0.0005 & 0.07 
                      & 84.12 & $<$ 0.0005 & 0.11 
                      & 259.29 & $<$ 0.001 & 0.19 \\
    Age            & 31.71 & $<$ 0.005 & 0.07 
                   & 88.67 & $<$ 0.01  & 0.10 
                   & 322.27 & $<$ 0.005 & 0.19 \\
    \hdashline
    \addlinespace[0.5ex]
    halting        & 106.82  & $<$ 0.0001  & 0.22
                   & 180.88  & $<$ 0.0001  & 0.28 
                   & 232.32  & $<$ 0.0001  & 0.32 \\
    fluent         & 89.58   & $<$ 0.0001  & 0.20 
                   & 111.39  & $<$ 0.0001  & 0.22 
                   & 185.69  & $<$ 0.0001  & 0.28 \\
    intellectual   & 47.48   & $<$ 0.0001  & 0.14
                   & 72.81   & $<$ 0.0001  & 0.18 
                   & 113.55  & $<$ 0.0005  & 0.22 \\
    sharp          & 35.86   & $<$ 0.0001  & 0.12 
                   & 36.74   & $<$ 0.005   & 0.13 
                   & 106.33  & $<$ 0.001   & 0.22 \\
    clear          & 31.28   & $<$ 0.0001  & 0.12 
                   & 36.89   & $<$ 0.005   & 0.13 
                   & 84.97   & $<$ 0.05    & 0.19 \\
    kind           & 28.53   & $<$ 0.0001  & 0.11 
                   & 32.87   & $<$ 0.005   & 0.12 
                   & 103.90  & $<$ 0.001   & 0.21 \\
    tensed         & 26.00   & $<$ 0.0001  & 0.11 
                   & 39.92   & $<$ 0.0005  & 0.13 
                   & 94.49   & $<$ 0.01    & 0.20 \\
    modest         & 21.17   & $<$ 0.0001  & 0.10 
                   & 25.84   & $<$ 0.05    & 0.11 
                   & 113.35  & $<$ 0.0005  & 0.22 \\
    reassuring     & 17.96   & $<$ 0.0005  & 0.09 
                   & 32.00   & $<$ 0.01    & 0.12 
                   & 94.42   & $<$ 0.01    & 0.20 \\
    weak           & 14.66   & $<$ 0.005   & 0.08 
                   & 27.37   & $<$ 0.05    & 0.11 
                   & 88.19   & $<$ 0.05    & 0.20 \\
    soft           & 13.93   & $<$ 0.005   & 0.08 
                   & 25.72   & $<$ 0.05    & 0.11 
                   & --      & --          & --   \\
    intense        & 11.76   & $<$ 0.01    & 0.07 
                   & --      & --          & --   
                   & 108.96  & $<$ 0.0005  & 0.22 \\
    relaxed        & 11.40   & $<$ 0.01    & 0.07 
                   & 27.14   & $<$ 0.05    & 0.11 
                   & --      & --          & --   \\
    mature         & 10.40   & $<$ 0.05    & 0.07 
                   & 31.59   & $<$ 0.01    & 0.12 
                   & 93.09   & $<$ 0.01    & 0.20 \\
    wild           & 10.20   & $<$ 0.05    & 0.07 
                   & 34.26   & $<$ 0.005   & 0.12 
                   & 119.72  & $<$ 0.0001  & 0.23 \\
    elegant        & 9.64    & $<$ 0.05    & 0.06 
                   & 42.10   & $<$ 0.0005  & 0.14 
                   & 119.88  & $<$ 0.0001  & 0.23 \\
    refreshing     & 9.34    & $<$ 0.05    & 0.06 
                   & 31.10   & $<$ 0.01    & 0.12 
                   & 86.43   & $<$ 0.05    & 0.19 \\
    raspy          & 8.66    & $<$ 0.05    & 0.06 
                   & --      & --          & --   
                   & --      & --          & --   \\
    strict         & 8.43    & $<$ 0.05    & 0.06
                   & --      & --          & --   
                   & --      & --          & --   \\
    sexy           & 7.94    & $<$ 0.05    & 0.06 
                   & --      & --          & --   
                   & --      & --          & --   \\
    cool           & --      & --          & --
                   & 33.74   & $<$ 0.005   & 0.12 
                   & --      & --          & --   \\
    hard           & --      & --          & --
                   & 26.25   & $<$ 0.05    & 0.11 
                   & --      & --          & --   \\
    light          & --      & --          & --
                   & 26.25   & $<$ 0.05    & 0.11 
                   & --      & --          & --   \\
    muffled        & --      & --          & --
                   & 25.20   & $<$ 0.05    & 0.10 
                   & --      & --          & --   \\
    cute           & --      & --          & --
                   & --      & --          & --
                   & 87.74   & $<$ 0.05    & 0.20 \\
    bright         & --      & --          & --
                   & --      & --          & --
                   & 82.59   & $<$ 0.05    & 0.19 \\
    \bottomrule
  \end{tabular}
\end{table}

\begin{table}
  \caption{Results of chi-squared tests (PLM: MP BERT).}
  \label{tab: chi mpbert}
  \centering
  \small
  \begin{tabular}{l r l r r l r r l r}
    \toprule
    \multirow{2}{*}{Characteristic} & \multicolumn{3}{c}{$k=4$} & \multicolumn{3}{c}{$k=16$} & \multicolumn{3}{c}{$k=64$}\\
    \cmidrule(lr){2-4} \cmidrule(lr){5-7} \cmidrule(lr){8-10} & $\chi^2$ & $p$ & $V$ & $\chi^2$ & $p$ & $V$ & $\chi^2$ & $p$ & $V$ \\
    \midrule
    Gender expression & 35.25 & $<$ 0.0001 & 0.07 
                      & 98.76 & $<$ 0.0001 & 0.12
                      & 325.15 & $<$ 0.0001 & 0.22 \\
    Age            & 29.97 & $<$ 0.005 & 0.07 
                   & 83.81 & $<$ 0.05  & 0.10 
                   & 319.45 & $<$ 0.005 & 0.19 \\
    \hdashline
    \addlinespace[0.5ex]
    halting         & 127.19   & $<$ 0.0001  & 0.24
                    & 168.90   & $<$ 0.0001  & 0.27 
                    & 231.57   & $<$ 0.0001  & 0.32 \\
    fluent          & 97.46    & $<$ 0.0001  & 0.21 
                    & 118.81   & $<$ 0.0001  & 0.23 
                    & 166.80   & $<$ 0.0001  & 0.27 \\
    intellectual    & 46.47    & $<$ 0.0001  & 0.14 
                    & 73.42    & $<$ 0.0001  & 0.18 
                    & 153.35   & $<$ 0.0001  & 0.26 \\
    sharp           & 43.69    & $<$ 0.0001  & 0.14 
                    & 51.50    & $<$ 0.0001  & 0.15 
                    & 105.55   & $<$ 0.001   & 0.21 \\
    kind            & 33.25    & $<$ 0.0001  & 0.12 
                    & 58.54    & $<$ 0.0001  & 0.16 
                    & 98.39    & $<$ 0.005   & 0.21 \\
    clear           & 26.16    & $<$ 0.0001  & 0.11 
                    & 47.61    & $<$ 0.0001  & 0.14 
                    & 101.96   & $<$ 0.005   & 0.21 \\
    weak            & 20.88    & $<$ 0.0005  & 0.10 
                    & 44.65    & $<$ 0.0005  & 0.14 
                    & --       & --          & --   \\
    modest          & 19.76    & $<$ 0.0005  & 0.09 
                    & 44.65    & $<$ 0.0001  & 0.14 
                    & 83.65    & $<$ 0.05    & 0.19 \\
    tensed          & 18.84    & $<$ 0.0005  & 0.09 
                    & 42.46    & $<$ 0.0005  & 0.14 
                    & 87.17    & $<$ 0.05    & 0.19 \\
    refreshing      & 18.30    & $<$ 0.0005  & 0.09
                    & 26.48    & $<$ 0.05    & 0.11 
                    & --       & --          & --   \\
    cool            & 16.36    & $<$ 0.001   & 0.08 
                    & 29.73    & $<$ 0.05    & 0.11 
                    & --       & --          & --   \\
    mature          & 15.98    & $<$ 0.005   & 0.08 
                    & 29.37    & $<$ 0.05    & 0.11 
                    & 87.31    & $<$ 0.05    & 0.19 \\
    reassuring      & 14.42    & $<$ 0.005   & 0.08 
                    & --       & --          & --   
                    & --       & --          & --   \\
    soft            & 13.18    & $<$ 0.005   & 0.08 
                    & 29.45    & $<$ 0.05    & 0.11 
                    & --       & --          & --   \\
    relaxed         & 12.09    & $<$ 0.01    & 0.07 
                    & 32.26    & $<$ 0.01    & 0.12 
                    & --       & --          & --   \\
    elegant         & 11.28    & $<$ 0.05    & 0.07 
                    & 48.02    & $<$ 0.0001  & 0.14 
                    & 115.96   & $<$ 0.0001  & 0.22 \\
    hard            & 11.17    & $<$ 0.05    & 0.07 
                    & 25.11    & $<$ 0.05    & 0.10 
                    & --       & --          & --   \\
    raspy           & 10.00    & $<$ 0.05    & 0.07 
                    & --       & --          & --   
                    & --       & --          & --   \\
    unique          & 8.87     & $<$ 0.05    & 0.06 
                    & --       & --          & --   
                    & --       & --          & --   \\
    muffled         & 8.69     & $<$ 0.05    & 0.06 
                    & 31.50    & $<$ 0.01    & 0.12 
                    & --       & --          & --   \\
    sexy            & 8.03     & $<$ 0.05    & 0.06 
                    & --       & --          & --   
                    & --       & --          & --   \\
    intense         & 7.86     & $<$ 0.05    & 0.06 
                    & --       & --          & --   
                    & 82.94    & $<$ 0.05    & 0.19 \\
    friendly        & 7.83     & $<$ 0.05    & 0.06 
                    & 37.44    & $<$ 0.005   & 0.13 
                    & 94.16    & $<$ 0.01    & 0.20 \\
    wild            & --       & --          & -- 
                    & 36.32    & $<$ 0.005   & 0.13 
                    & 137.45   & $<$ 0.0001  & 0.24 \\
    bright          & --       & --          & -- 
                    & 30.93    & $<$ 0.01    & 0.12 
                    & 84.10    & $<$ 0.05    & 0.19 \\
    dark            & --       & --          & -- 
                    & --       & --          & -- 
                    & 85.95    & $<$ 0.05    & 0.19 \\
    \bottomrule
  \end{tabular}
\end{table}

\subsubsection{Chi-squared test}

We conducted chi-squared tests across all speakers to investigate the association between the speaker embedding clusters and each speaker characteristic annotated in LibriTTS-P. We first removed degree adverbs such as ``slightly'' and ``very'' in the annotations. Since nearly all speakers were annotated with gender expression and age, we categorized speakers into \textit{feminine}, \textit{masculine}, \textit{gender-neutral}, and \textit{unlabeled} for gender expression, and into \textit{young}, \textit{adult-like}, \textit{middle-aged}, \textit{old}, and \textit{unlabeled} for age. Accordingly, the chi-squared tests for the two attributes were performed on multi-class contingency tables. For all other characteristics, we performed binary classification based on whether the given characteristic was present in the speaker's annotation.

We set the number of $k$-means clusters $k$ to 4, 16, and 64. For each setting, we report the $\chi^2$ statistic, $p$-value, and Cram\'er's $V$, and present the results where the $p$-value is less than 0.05. Let $\mathcal{S}$ denote the resulting list of statistically significant characteristics ($p<0.05$). The results are shown in Table~\ref{tab: chi resnet}, Table~\ref{tab: chi bert}, and Table~\ref{tab: chi mpbert}, where the entries are sorted in descending order of $V$ for $k=4$ to facilitate comparisons. We observe the following trends:
\begin{itemize}
    \item The two types of phrasing speaker embedding clusters exhibit similar behavior in the tests: as $k$ increases, both $\chi^2$ and $V$ generally increase, while the size of $\mathcal{S}$ tends to decrease. In contrast, the trend differs for the ResNet-TDNN embeddings: the size of $\mathcal{S}$ increases as $k$ rises from 4 to 16, and then remains comparatively stable as $k$ increases further to 64. This suggests that supervised speaker embeddings may retain more fine-grained speaker-related information that aligns well with the annotated characteristics.

    \item For all types of embedding clusters, gender expression and age consistently show relatively high Cram\'er’s $V$ values. This indicates a potential association between the two speaker attributes and RP insertion behavior.

    \item Notably, \textit{halting} and \textit{fluent} consistently show the highest Cram\'er's $V$ values in the phrasing speaker embedding clusters. Since both relate to speech fluency, this result is intuitive and suggests that speech fluency is closely related to RP insertion frequency. In contrast, for the ResNet-TDNN embeddings, the characteristics with the highest Cramér’s $V$ values are \textit{wild} and \textit{elegant}, which reflect more stylistic or personality-related characteristics.

    \item Other characteristics such as \textit{intellectual}, \textit{sharp}, \textit{clear}, \textit{kind} also exhibit relatively high Cram\'er's $V$ values  in the phrasing speaker embedding clusters. These characteristics may reflect speaking rate or articulation clarity, both of which can influence RP insertion.

    \item In terms of the association with speaker characteristics, the two phrasing speaker embedding clusters exhibit similar behavior. Although the rankings of $V$ differ between the phrasing embedding clusters and the ResNet-TDNN embedding clusters, $\mathcal{S}$ of ResNet-TDNN embedding clusters largely cover that of phrasing speaker embedding clusters. This provides objective support for the feasibility of incorporating PSVM embeddings into phrasing models for few-shot adaptation.
\end{itemize}

Since the phrasing speaker embeddings used for clustering were trained solely on text and RP positions, the observed correlations between the clusters and certain speaker characteristics suggest that the phrasing models can capture and encode phrasing-related characteristics through the training objective. Furthermore, the results of the chi-squared tests provide an objective basis for identifying speaker characteristics that are most likely to influence phrasing behavior.

\section{Conclusions}

This work confirmed the effectiveness of incorporating speaker embeddings and phoneme-level PLMs into the phrasing model for TTS systems. The observed improvements in phrasing accuracy and MOS affirm that different speakers have distinct RP insertion styles and that phrasing is related to phoneme-level features. We also proposed a few-shot adaptation method, demonstrating that speaker embeddings from PSVMs can generalize our proposed model to unseen speakers without fine-tuning.

\section{Limitations and future work}

The exclusive use of training data from the LibriTTS-R corpus, which comprises only audiobook readings, limits the generalization of our phrasing models to speech utterances with varying styles and contexts. To address this, expanding the training dataset to include a more diverse range of speech types could not only improve the models' robustness but also enable few-shot or even zero-shot learning capabilities for out-of-domain speakers.

Additionally, while PSVM embeddings capture the speaker's prosody features, phrasing is also influenced by the prosody of each utterance. Developing a method to model and predict utterance-level prosody could allow for the integration of speaker-specific and utterance-level features, potentially further enhancing phrasing accuracy.

Lastly, while phoneme-level PLMs contribute to improved performance in phrasing models, a key limitation lies in our inability to disentangle the respective effects of phonemes and subwords on RP position prediction. This entanglement complicates the formulation of a theoretical explanation for the observed improvements. A deeper investigation into the roles of phoneme representations remains an important direction for future work.

\section{Acknowledgements}
Part of this research and development work was supported by JSPS KAKENHI 22K17945, JST SPRING JSMJSP2108, and JST Moonshot Grant Number JPMJMS2237.

\appendix

\section{Implementation details and improvements of Matcha-TTS}
\label{appendix a}

Since the original Matcha-TTS model introduced in~\cite{matcha-tts} is designed for LJSpeech~\cite{ljspeech} and VCTK~\cite{vctk}, which are small speech corpora, we made some improvements to adapt it to our TTS task with using a larger multi-speaker dataset:
\begin{itemize}
    \item We replaced the pre-trained HiFi-GAN~\cite{kong20hifigan} with the pre-trained Vocos~\cite{vocos} as the vocoder, which yields better performance.
    
    \item We replaced the \textit{duration predictor} with the one in VITS~\cite{kim21vits} \footnote{\url{https://github.com/jaywalnut310/vits/blob/2e561ba58618d021b5b8323d3765880f7e0ecfdb/models.py#L98}}. Structurally, the \textit{duration predictor} in Matcha-TTS derives from that of VITS but removes the speaker embedding component. This removal notably degrades duration prediction performance when the dataset contains a large number of speakers. The integration of the speaker embedding and hidden state was modified from element-wise addition to concatenation, which aligns better with the coding conventions of Matcha-TTS.
    
    \item During training, we used a learning rate scheduler as described in~\ref{sec: train tts} and reduced the \textit{gradient\_clip\_val} to 1.0 to avoid gradient explosion. In the data loading, we also followed the same strategy as VITS by discarding samples with phoneme lengths exceeding 190\footnote{\url{https://github.com/jaywalnut310/vits/blob/2e561ba58618d021b5b8323d3765880f7e0ecfdb/data_utils.py#L34}}, which ensured training efficiency.

    \item Finally, we increased the sizes of model parameters as summarized in Table~\ref{tab: matcha-tts parameter}.
\end{itemize}

\begin{table*}[t]
  \caption{List of increased parameter sizes in Matcha-TTS.}
  \label{tab: matcha-tts parameter}
  \centering
  \small
  \begin{tabular}{l l r r}
    \toprule
    Component & Parameter & Original size & Increased size \\
    \toprule
    \multirow{5}{*}{Encoder} & n\_channels & 192 & 384 \\
    & filter\_channels & 768 & 2,048 \\
    & filter\_channels\_dp & 256 & 1,024 \\
    & n\_heads & 2 & 4 \\
    \midrule
    \multirow{3}{*}{Decoder} & n\_blocks & 1 & 4 \\
    & num\_mid\_blocks & 2 & 8 \\
    & num\_heads & 2 & 8 \\
    \bottomrule
  \end{tabular}
\end{table*}

\section{Related algorithms of the embedding adapter}
\label{appendix b}

\begin{algorithm}[H]
\caption{Training and inference procedures of the embedding adapter}
\textbf{Training procedure} \\
\textbf{Input:} the training set $\{\mathbf{e}_m, \mathbf{e}'_m\}_{m \in \mathcal{M}}$, the parameters of the embedding adapter $\boldsymbol{\theta}$.
\begin{algorithmic}[1]
\Repeat
    \State Sample a mini-batch $\mathcal{B} \subset \mathcal{M}, 
    \mathbf{E}_\mathcal{B} = \{\mathbf{e}_b\}_{b \in \mathcal{B}}, 
    \mathbf{E}'_\mathcal{B} = \{\mathbf{e}'_b\}_{b \in \mathcal{B}}$
    \State Generate the output of the adapter $\hat{\mathbf{E}}'_\mathcal{B} = \{\boldsymbol{f_\theta}(\mathbf{e}_b)\}_{b \in \mathcal{B}}$
    \State Apply gradient descent to minimize $\mathcal{L}(\boldsymbol{\theta})
    = \mathbb{E}||\hat{\mathbf{E}}'_\mathcal{B} - \mathbf{E}'_\mathcal{B}||^2.$
\Until{convergence}
\end{algorithmic}
\textbf{Inference procedure} \\
\textbf{Input:} $\{\mathbf{e}_k\}_{k \in \mathcal{K}}$
\begin{algorithmic}[1]
\For{each $k \in \mathcal{K}$}
    \State $\hat{\mathbf{e}}'_k = \boldsymbol{f_\theta}(\mathbf{e}_k)$
\EndFor
\end{algorithmic}
\textbf{Output:} $\{\hat{\mathbf{e}}'_k\}_{k \in \mathcal{K}}$
\end{algorithm}

\section{Resources of pre-trained models}
\label{appendix c}

\begin{table*}[t]
  \caption{Link list of pre-trained models.}
  \label{tab: resources}
  \centering
  \small
  \begin{tabular}{l l >{\raggedright\arraybackslash}p{10cm}}
    \toprule
    Model & Version & Link \\
    \midrule
    \multirow{2}{*}{BERT} & BASE & \url{https://huggingface.co/google-bert/bert-base-uncased}\\
    & LARGE & \url{https://huggingface.co/google-bert/bert-large-uncased}\\
    \multirow{2}{*}{XLNET} & BASE & \url{https://huggingface.co/xlnet/xlnet-base-cased}\\
    & LARGE & \url{https://huggingface.co/xlnet/xlnet-large-cased}\\
    \multirow{2}{*}{RoBERTa} & BASE & \url{https://huggingface.co/FacebookAI/roberta-base}\\
    & LARGE & \url{https://huggingface.co/FacebookAI/roberta-large}\\
    \multirow{2}{*}{ALBERT} & BASE & \url{https://huggingface.co/albert/albert-base-v2}\\
    & LARGE & \url{https://huggingface.co/albert/albert-large-v2}\\
    \multirow{2}{*}{DeBERTaV3} & BASE & \url{https://huggingface.co/microsoft/deberta-v3-base}\\
    & LARGE & \url{https://huggingface.co/microsoft/deberta-v3-large}\\
    \midrule
    MP BERT & -- & \url{https://huggingface.co/ydqmkkx/mpbert}\\
    PL BERT & -- & \url{https://huggingface.co/ydqmkkx/plbert}\\
    \midrule
    ECAPA-TDNN & -- & \url{https://huggingface.co/speechbrain/spkrec-ecapa-voxceleb}\\
    ResNet-TDNN & -- & \url{https://huggingface.co/speechbrain/spkrec-resnet-voxceleb}\\
    SpeakerNet & -- & \url{https://catalog.ngc.nvidia.com/orgs/nvidia/teams/nemo/models/speakerverification_speakernet}\\
    TitaNet & LARGE & \url{https://catalog.ngc.nvidia.com/orgs/nvidia/teams/nemo/models/titanet_large}\\
    \bottomrule
  \end{tabular}
\end{table*}

We provide links to the pre-trained models used in this work in Table~\ref{tab: resources}. Among them, MP BERT and PL BERT are our implementations, and ECAPA-TDNN and ResNet-TDNN are available via the SpeechBrain toolkit~\cite{ravanelli21speechbrain}.

\section{The impact of training data difference between PSVMs and TTS models}
\label{appendix d}

Given that \texttt{SpeechBrain} provides complete training pipelines for ECAPA-TDNN and ResNet-TDNN, while \texttt{NeMo} does not release training code for SpeakerNet nor TitaNet, we focus our exploration of the impact of training data inconsistency on ECAPA-TDNN and ResNet-TDNN. Specifically, we first followed the official configurations and pipelines to train ECAPA-TDNN and ResNet-TDNN, with the training and validation sets as shown in Table~\ref{tab: statistics}. Then,  based on these two PSVMs, we train phrasing models with BERT\textsubscript{BASE} and MP BERT as the PLMs. The phrasing results for seen and unseen speakers are presented in Table~\ref{tab: seen-consistency} and Table~\ref{tab: unseen-consistency}, respectively. The arrows in the tables indicate the relative changes in F\textsubscript{0.5} scores after ensuring training data consistency between the PSVMs and the phrasing models, compared to previous results.

As shown by the arrows, the impact of training data inconsistency between the PSVM and the phrasing models varies across different tasks. Its effect can be either positive or negative, depending on both PSVMs and PLMs. Therefore, we are unable to draw a universal conclusion or assert that the impact is consistently beneficial or detrimental.

However, we observe that when training on LibriTTS-R, the ECAPA-TDNN with BERT\textsubscript{BASE} leads to a substantial performance drop on the test-unseen set. Besides, from a practical standpoint, using the PSVMs officially pre-trained on the Voxceleb corpus is more convenient and generally preferable for training phrasing models.

\begin{table*}[t]
  \caption{Phrasing results for seen speakers. The training and validation sets of the phrasing models and PSVMs are the same.}
  \label{tab: seen-consistency}
  \centering
  \scalebox{0.9}{
  \begin{tabular}{l l c c c c c c c}
    \toprule
    \multirow{2}{*}{PLM} & \multirow{2}{*}{Speaker embeddings} & \multicolumn{3}{c}{Frozen} & \multicolumn{3}{c}{Trainable}\\ \cmidrule(lr){3-5} \cmidrule(lr){6-9} & & Precision & Recall & \textbf{F\textsubscript{0.5}} & Precision & Recall & \textbf{F\textsubscript{0.5}}\\
    \toprule
    \multirow{2}{*}{BERT\textsubscript{BASE}} & ECAPA-TDNN & 0.2869 & 0.2312 & 0.2737 ($\downarrow$) & 0.3433 & 0.2364 & 0.3148 ($\downarrow$)\\
    & ResNet-TDNN & 0.5864 & 0.2585 & 0.4678 ($\uparrow$) & 0.5828 & 0.2674 & 0.4715 ($\uparrow$)\\
    \midrule
    \multirow{2}{*}{MP BERT} & ECAPA-TDNN & 0.5802 & 0.2835 & 0.4798 ($\uparrow$) & 0.5826 & 0.2920 & 0.4859 ($\uparrow$)\\
    & ResNet-TDNN & 0.5922 & 0.3012 & 0.4963 ($\uparrow$) & 0.5893 & 0.3085 & 0.4986 ($\downarrow$)\\
    \bottomrule
  \end{tabular}
  }
\end{table*}

\begin{table*}[t]
  \caption{Phrasing results of few-shot adaptation for unseen speakers. The training and validation sets of the phrasing models and PSVMs are the same.}
  \label{tab: unseen-consistency}
  \centering
  \small
  \begin{subtable}{1.0\linewidth}
    \caption{\footnotesize Proposed models with frozen speaker embeddings.}
    \centering
    \small
      \begin{tabular}{l l c c c c c c c c}
        \toprule
        \multirow{2}{*}{PLM} & \multirow{2}{*}{Speaker embedding} &\multicolumn{7}{c}{Samples} \\ \cmidrule(lr){3-9} & & 1 & 5 & 10 & 20 & 30 & 40 & 50\\
        \toprule
        \multirow{2}{*}{BERT\textsubscript{BASE}}
        & ECAPA-TDNN & 0.1844 ($\downarrow$) & 0.1522 ($\downarrow$) & 0.1800 ($\downarrow$) & 0.1744 ($\downarrow$) & 0.1694 ($\downarrow$) & 0.1620 ($\downarrow$) & 0.1613 ($\downarrow$)\\
        & ResNet-TDNN & 0.2938 ($\uparrow$) & 0.3112 ($\downarrow$) & 0.3254 ($\downarrow$) & 0.3272 ($\downarrow$) & 0.3340 ($\downarrow$) & 0.3318 ($\downarrow$) & 0.3319 ($\downarrow$)\\
        \midrule
        \multirow{2}{*}{MP BERT} 
        & ECAPA-TDNN & 0.3072 ($\uparrow$) & 0.3336 ($\uparrow$) & 0.3402 ($\uparrow$) & 0.3425 ($\downarrow$) & 0.3452 ($\downarrow$) & 0.3487 ($\downarrow$) & 0.3458 ($\downarrow$)\\
        & ResNet-TDNN & 0.3056 ($\uparrow$) & 0.3235 ($\downarrow$) & 0.3257 ($\downarrow$) & 0.3259 ($\downarrow$) & 0.3259 ($\downarrow$) & 0.3315 ($\downarrow$) & 0.3282 ($\downarrow$)\\
        \bottomrule
      \end{tabular}
    \end{subtable}
    \vspace{5pt} 
    
  \begin{subtable}{1.0\linewidth}
    \caption{\footnotesize Proposed models with trainable speaker embeddings.}
    \centering
    \small
      \begin{tabular}{l l c c c c c c c c}
        \toprule
        \multirow{2}{*}{PLM} & \multirow{2}{*}{Speaker embedding} &\multicolumn{7}{c}{Samples} \\ \cmidrule(lr){3-9} & & 1 & 5 & 10 & 20 & 30 & 40 & 50\\
        \toprule
        \multirow{2}{*}{BERT\textsubscript{BASE}}
        & ECAPA-TDNN & 0.0767 ($\downarrow$) & 0.0912 ($\downarrow$) & 0.0890 ($\downarrow$) & 0.0794 ($\downarrow$) & 0.0908 ($\downarrow$) & 0.0881 ($\downarrow$) & 0.0885 ($\downarrow$)\\
        & ResNet-TDNN & 0.3161 ($\uparrow$) & 0.3506 ($\uparrow$) & 0.3523 ($\downarrow$) & 0.3560 ($\downarrow$) & 0.3645 ($\downarrow$) & 0.3686 ($\downarrow$) & 0.3618 ($\downarrow$)\\
        \midrule
        \multirow{2}{*}{MP BERT} 
        & ECAPA-TDNN & 0.3183 ($\uparrow$) & 0.3493 ($\uparrow$) & 0.3472 ($\uparrow$) & 0.3549 ($\uparrow$) & 0.3591 ($\uparrow$) & 0.3600 ($\uparrow$) & 0.3582 ($\uparrow$)\\
        & ResNet-TDNN & 0.3283 ($\uparrow$) & 0.3405 ($\uparrow$) & 0.3412 ($\downarrow$) & 0.3480 ($\downarrow$) & 0.3512 ($\downarrow$) & 0.3464 ($\downarrow$) & 0.3455 ($\downarrow$)\\
        \bottomrule
      \end{tabular}
    \end{subtable}
\end{table*}

\section{Phrasing models using MP BERT}
\label{appendix e}

\begin{table*}[t]
  \caption{Phrasing results of the phrasing models with MP BERT for seen speakers.}
  \label{tab: mpbert}
  \centering
  \small

  \begin{subtable}{1.0\linewidth}
    \caption{\footnotesize Baseline model.}
    \centering
    \small
      \begin{tabular}{c c c}
        \toprule
        Precision & Recall & \textbf{F\textsubscript{0.5}}\\
        \midrule
        0.4957 & 0.2410 & 0.4092\\
        \bottomrule
      \end{tabular}
  \end{subtable}
  \vspace{5pt} 
  
  \begin{subtable}{1.0\linewidth}
    \caption{\footnotesize Proposed models with various PSVM embeddings.}
    \centering
    \small
      \begin{tabular}{l l l c c c c c c c c c c}
        \toprule
        \multirow{2}{*}{Speaker embeddings} & \multicolumn{3}{c}{Frozen} & \multicolumn{3}{c}{Trainable}\\ \cmidrule(lr){2-4} \cmidrule(lr){5-7} & Precision & Recall & \textbf{F\textsubscript{0.5}} & Precision & Recall & \textbf{F\textsubscript{0.5}}\\
        \midrule
        Xavier initialization & -- & -- & -- & 0.6238 & 0.2773 & 0.4991\\
        ECAPA-TDNN & 0.5579 & 0.2554 & 0.4511 & 0.5457 & 0.2954 & 0.4666\\
        ResNet-TDNN & 0.5970 & 0.2831 & \textbf{0.4886} & 0.6053 & 0.2965 & 0.5010\\
        SpeakerNet & 0.5535 & 0.2920 & 0.4695 & 0.6055 & 0.2997 & \underline{\textbf{0.5029}}\\
        Titanet\textsubscript{LARGE} & 0.5537 & 0.2806 & 0.4635 & 0.5949 & 0.3061 & 0.5004\\
        \bottomrule
      \end{tabular}
    \end{subtable}
\end{table*}

To further investigate the effectiveness of MP BERT, we applied it to both the baseline model and the proposed model with various PSVM embeddings. After training, their results on the test-seen set are shown in Table~\ref{tab: mpbert}:
\begin{itemize}
    \item Compared to the results in Table~\ref{tab: baseline seen results} and Table~\ref{tab: spk}, MP BERT significantly improved the F\textsubscript{0.5} score for both the baseline and proposed model over BERT\textsubscript{BASE}, highlighting its effectiveness.

    \item Among the proposed models, the highest phrasing accuracy was achieved using trainable SpeakerNet embeddings. However, the improvement over randomly initialized embeddings was minimal. This suggests that for seen speakers, randomly initialized embeddings remain a convenient and effective option.
\end{itemize}

\section{Comparisons of different sets on RP distributions}
\label{appendix f}

\begin{table*}[t]
  \caption{Statistical comparison of RP distributions across different sets.}
  \label{tab: stat rp sets}
  \centering
  \small
  \begin{subtable}{1.0\linewidth}
    \centering
    \caption{\footnotesize RP insertion frequency ($f_{RP}$) across speakers.}
      \begin{tabular}{l c c c}
        \toprule
        Set & mean $\pm$ std & $W_1$ & KS\\
        \toprule
        Training & 0.1007 $\pm$ 0.0690 & -- & --\\
        Validation-seen & 0.0978 $\pm$ 0.0766 & 0.0075 & 0.0686\\
        Test-seen & 0.0998 $\pm$ 0.0791 & 0.0079 & 0.0663\\
        Validation-unseen & 0.0604 $\pm$ 0.0505 & 0.0404 & 0.3291\\
        Test-unseen & 0.0587 $\pm$ 0.0465 & 0.0422 & 0.3222\\
        \bottomrule
      \end{tabular}
  \end{subtable}
  \vspace{5pt} 

  \begin{subtable}{1.0\linewidth}
    \centering
    \caption{\footnotesize Average RP duration ($\overline{T}_{RP}$) across speakers.}
      \begin{tabular}{l c c c}
        \toprule
        Set & mean $\pm$ std & $W_1$ & KS\\
        \toprule
        Training & 0.1885 $\pm$ 0.0527 & -- & --\\
        Validation-seen & 0.1857 $\pm$ 0.0763 & 0.0174 & 0.1204\\
        Test-seen & 0.1901 $\pm$ 0.0849 & 0.0189 & 0.1080\\
        Validation-unseen & 0.1850 $\pm$ 0.0580 & 0.0077 & 0.1187\\
        Test-unseen & 0.1917 $\pm$ 0.0505 & 0.0095 & 0.1030\\
        \bottomrule
      \end{tabular}
  \end{subtable}
\end{table*}

\begin{figure*}[h]
    \centering
    \includegraphics[width=1.0\linewidth, clip]{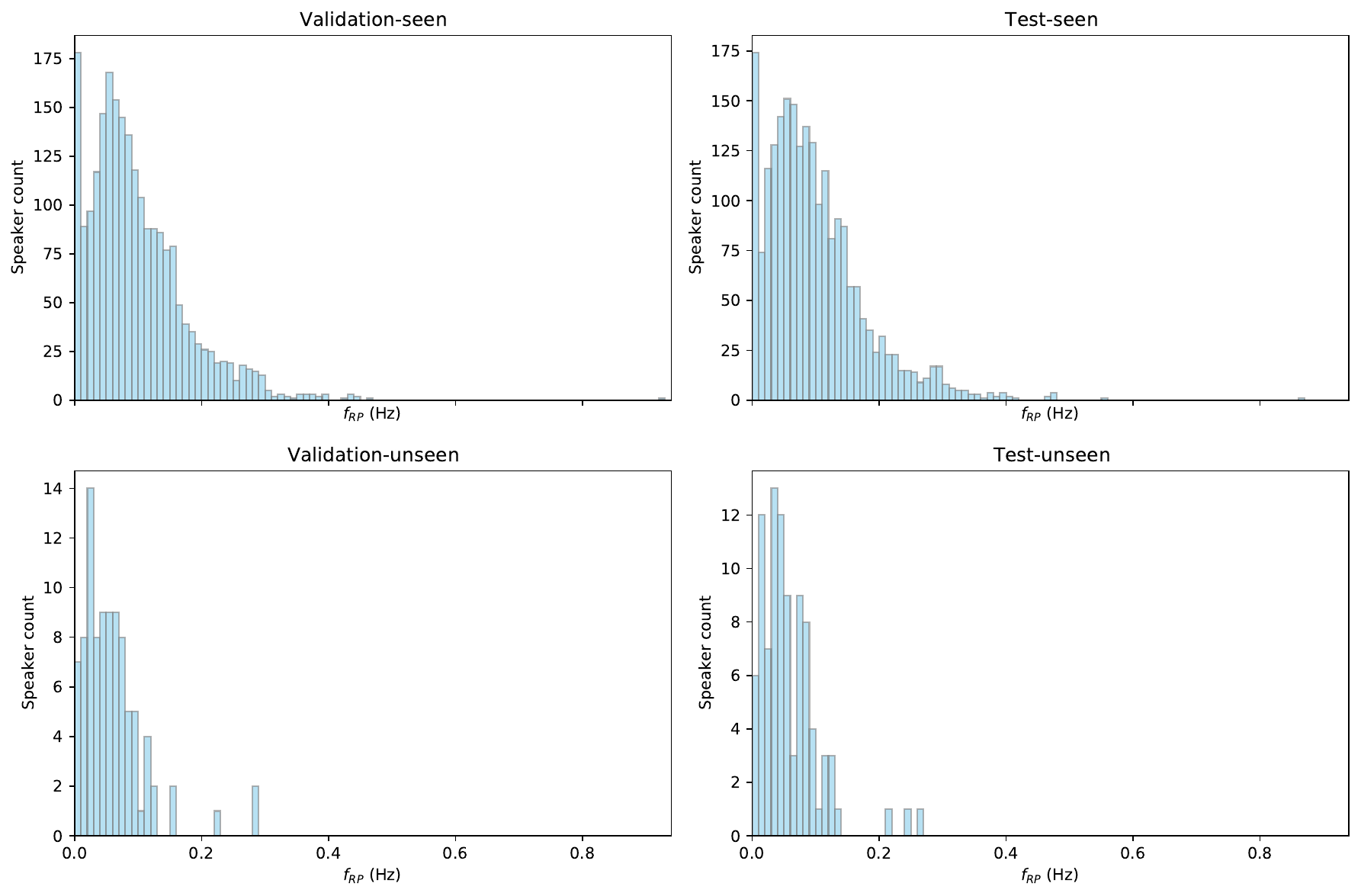}
    \caption{Histograms of RP insertion frequency ($f_{RP}$) across different sets.}
    \label{fig: overfit rp frequency}
\end{figure*}

To illustrate and quantify the differences in RP distributions across different sets, we adopt two statistical features introduced in Section~\ref{sec: rp distribution}: RP insertion frequency ($f_{RP}$) and average RP duration ($\overline{T}_{RP}$). As shown in Table~\ref{tab: stat rp sets}, we report the mean and standard deviation (std) for each set. In addition, we provide the 1-Wasserstein distance ($W_1$) and the two-sample Kolmogorov-Smirnov (KS) statistic with respect to the training set. From the table, we observe the following:
\begin{itemize}
    \item For $f_{RP}$, which is directly related to the phrasing task, the sets of unseen speakers exhibit substantial differences compared with the sets of seen speakers across all reported metrics. To visualize the difference more clearly, we present the distributions for each set in Fig.\ref{fig: overfit rp frequency}, which can be compared with Fig.\ref{fig: rp frequency}. Consequently, for phrasing models trained on the training set and tuned on the validation-seen set to determine thresholds, the strong performance of MP-BERT on the test-seen set fails to generalize to the test-unseen set.

    \item For $\overline{T}_{RP}$, which is only relevant to the TTS models in this work, the differences between the sets of unseen speakers and the training set are relatively small. Moreover, the sets of unseen speakers are even closer to the training set than the validation-seen and test-seen sets in terms of $W_1$. This observation partially explains why phrasing models using MP-BERT achieve comparable or even better performance than those using BERT\textsubscript{BASE} in subjective evaluations.
\end{itemize}

\clearpage
 \bibliographystyle{elsarticle-num} 
 \bibliography{cas-refs1}





\end{document}